\documentclass[twocolumn,english,onecolumn,pra,amsmath,amssymb,superscriptaddress,showpacs,longbibliography]{revtex4}
\usepackage[T1]{fontenc}
\usepackage[latin9]{inputenc}
\usepackage{CJK}
\usepackage{babel}
\usepackage{bm}
\usepackage{amsmath}
\usepackage{amssymb}
\usepackage{graphicx}
\usepackage{esint}
\usepackage[unicode=true,pdfusetitle,
 bookmarks=true,bookmarksnumbered=false,bookmarksopen=false,
 breaklinks=false,pdfborder={0 0 1},backref=false,colorlinks=false]
 {hyperref}
\hypersetup{
 colorlinks,citecolor=blue,linkcolor=blue,urlcolor=blue}

\makeatletter
\@ifundefined{textcolor}{}
{%
 \definecolor{BLACK}{gray}{0}
 \definecolor{WHITE}{gray}{1}
 \definecolor{RED}{rgb}{1,0,0}
 \definecolor{GREEN}{rgb}{0,1,0}
 \definecolor{BLUE}{rgb}{0,0,1}
 \definecolor{CYAN}{cmyk}{1,0,0,0}
 \definecolor{MAGENTA}{cmyk}{0,1,0,0}
 \definecolor{YELLOW}{cmyk}{0,0,1,0}
}

\usepackage{bm}
\usepackage{times}
\usepackage{xcolor}
\newcommand{\ket}[1]{\lvert #1 \rangle}

\makeatother

\begin{document}
\global\long\def\MXtwo{M\! X_{2}}
\global\long\def\kp{\bm{k}\cdot\bm{p}}

\begin{CJK}{GB}{}\global\long\def\vr{\bm{r}}
\global\long\def\vR{\bm{R}}
\global\long\def\vk{\bm{k}}
\global\long\def\vK{\bm{K}}

\global\long\def\bktwo#1#2{\langle#1|#2\rangle}

\global\long\def\bkthree#1#2#3{\langle#1|#2|#3\rangle}

\global\long\def\ket#1{|#1\rangle}
\global\long\def\bra#1{\langle#1|}

\global\long\def\ave#1{\langle#1\rangle}

\end{CJK}

\inputencoding{latin9}%
\selectlanguage{english}%

\title{Intervalley coupling by quantum dot confinement potentials in monolayer
transition metal dichalcogenides}

\date{\today}

\author{Gui-Bin Liu}

\affiliation{School of Physics, Beijing Institute of Technology, Beijing 100081,
China}

\affiliation{Department of Physics and Center of Theoretical and Computational
Physics, The University of Hong Kong, Hong Kong, China}

\author{Hongliang Pang}

\affiliation{Department of Physics and Center of Theoretical and Computational
Physics, The University of Hong Kong, Hong Kong, China}

\author{Yugui Yao}

\affiliation{School of Physics, Beijing Institute of Technology, Beijing 100081,
China}

\author{Wang Yao}

\thanks{wangyao@hku.hk}

\affiliation{Department of Physics and Center of Theoretical and Computational
Physics, The University of Hong Kong, Hong Kong, China}
\begin{abstract}
Monolayer transition metal dichalcogenides (TMDs) offer new opportunities
for realizing quantum dots (QDs) in the ultimate two-dimensional (2D)
limit. Given the rich control possibilities of electron valley pseudospin
discovered in the monolayers, this quantum degree of freedom can be
a promising carrier of information for potential quantum spintronics
exploiting single electrons in TMD QDs. An outstanding issue is to
identify the degree of valley hybridization, due to the QD confinement,
which may significantly change the valley physics in QDs from its
form in the 2D bulk. Here we perform a systematic study of the intervalley
coupling by QD confinement potentials on extended TMD monolayers.
We find that the intervalley coupling in such geometry is generically
weak due to the vanishing amplitude of the electron wavefunction at
the QD boundary, and hence valley hybridization shall be well quenched by the much
stronger spin-valley coupling in monolayer TMDs and the QDs can
well inherit the valley physics of the 2D bulk. We also discover sensitive dependence of
intervalley coupling strength on the central position and the lateral
length scales of the confinement potentials, which may possibly allow
tuning of intervalley coupling by external controls.
\end{abstract}


\maketitle

\section{Introduction}

Semiconductor quantum dots (QDs) have been widely explored in the
past several decades for technological applications such as lasers
and medical markers \cite{Tartakovskii_Tartakovskii_2012____Quantum}.
Spin and other quantum degrees of freedom of single electron or hole
confined in QDs are also extensively researched as potential carriers
of information for quantum computing and quantum spintronics \cite{Hanson_Vandersypen_2007_79_1217__Spins,Liu_Sham_2010_59_703__Quantum,Loss_DiVincenzo_1998_57_120__Quantum,Reimann_Manninen_2002_74_1283__Electronic,Burkard_Loss_1999_59_2070__Coupled}.
QDs in conventional semiconductors are realized either by forming
nanocrystals or by lateral confinements in two-dimensional (2D) heterostructures.
Such lateral confinements can be provided by the thickness variations
of the heterostructures or patterned electrodes \cite{Liu_Sham_2010_59_703__Quantum,Hanson_Vandersypen_2007_79_1217__Spins}.
The emergence of graphene \cite{CastroNeto_Geim_2009_81_109__electronic,Geim_Novoselov_2007_6_183__rise},
the 2D crystal with single atom thickness, has offered a new system
to host QDs. Because of the zero gap nature of graphene, QD confinement
has to be facilitated by terminations of the monolayer, and the various
geometries explored include the graphene nano-island connected to
source and drain contacts via narrow graphene constrictions \cite{Ponomarenko_Geim_2008_320_356__Chaotic,Guttinger_Ensslin_2009_103_46810__Electron,Wang_Chang_2011_99_112117__Gates},
and graphene nanoribbon of armchair edges with patterned electrodes
\cite{Trauzettel_Burkard_2007_3_192__Spin}. The properties of these
graphene QDs are affected significantly by the edges, and precise
control on edge terminations are thus desired. 

Monolayer group-VIB transition metal dichalcogenides (TMDs) are newly
emerged members of the 2D crystal family \cite{Wang_Strano_2012_7_699__Electronics}.
These TMDs have the chemical composition of MX$_{2}$ (M = Mo,W; X
= S,Se), and the monolayer has a structure of X-M-X covalently bonded
hexagonal quasi-2D network. The 2D bulk has a direct bandgap in the
visible frequency range \cite{splendiani_emerging_2010,mak_atomically_2010},
ideal for optoelectronic applications. This bandgap also makes possible
the realization of QDs without the assistance of termination edges
\cite{Kormanyos_Burkard_2014_4_11034__Spin}. QDs can be defined by
lateral confinement potentials on an extended monolayer, e.g. by patterned
electrodes, similar to the QDs in III-V heterostructures \cite{Hanson_Vandersypen_2007_79_1217__Spins}.
As the conduction and valence band edges are shifted in the same way
by the electrodes, the QDs will have a unique type-II band alignment
with its surroundings (cf. figure 1). Moreover, QD confinement potentials
can also be realized by the lateral heterojunctions between different
TMDs within a single uniform crystalline monolayer. Various band alignment
can form between different TMDs with different bandgaps and workfunctions.
Lateral heterostructures with MoSe$_{2}$ islands surrounded by WSe$_{2}$
on a crystalline monolayer has been realized very recently \cite{Huang_Xu_2014____Lateral}.
This can realize QD confinement, also with type-II band alignment,
for electrons at the MoSe$_{2}$ region. The band edge discontinuity
at the heterojunction, with an order of magnitude of 0.2--0.4 eV \cite{Rivera_Xu_2014___1403.4985_Observation,Lee_Kim_2014___1403.3062_Atomically,Furchi_Mueller_2014___1403.2652_Photovoltaic,Cheng_Duan_2014___1403.3447_Electroluminescence,Fang_Javey_2014___1403.3754_Strong},
forms the potential well (with a vertical wall). 

Besides the unique 2D geometries, monolayer TMD QDs are highly appealing
because of the interesting properties of the 2D bulk. The conduction
and valence band edges are both at the degenerate $K$ and $-K$ valleys
at the corners of the hexagonal Brillouin zone. Remarkably, the direct-gap
optical transitions are associated with a valley dependent selection
rule: left- (right-) handed circular polarized light excites interband
transitions in the $K$ ($-K$) valley only \cite{Yao_Niu_2008_77_235406__Valley,Xiao_Yao_2012_108_196802__Coupled}.
Based on this selection rule, optical pumping of valley polarization
\cite{zeng_valley_2012,mak_control_2012,cao_valley_selective_2012},
and optical generation of valley coherence \cite{Jones_Xu_2013_8_634__Optical},
have been demonstrated. Another remarkable property of the 2D TMDs
is the strong interplay between spin and the valley pseudospin \cite{Gong_Yao_2013___1303.3932_Magnetoelectric,Jones_Xu_2014_10_130__Spin,Xu_Heinz_2014____}.
These quantum degrees of freedom with versatile controllability well
suggest that single electron in monolayer TMD QDs can be promising
carrier of information for quantum spintronics, provided that the
bulk properties of interest can be inherited by the QDs. Since these
interesting bulk properties are associated with the valley pseudospin,
a central problem is whether valley is still a good quantum number
and whether the bulk valley physics is preserved in the QDs. A natural
concern is that the lateral QD confinement may result in valley hybridization,
like in silicon QDs \cite{Saraiva_Koiller_2009_80_81305__Physical,Koiller_Das_2004_70_115207__Shallow,Goswami_Eriksson_2007_3_41__Controllable,Friesen_Coppersmith_2010_81_115324__Theory,Culcer_Das_2010_82_155312__Quantum,Friesen_Coppersmith_2007_75_115318__Valley,Nestoklon_Ivchenko_2006_73_235334__Spin,Boykin_Lee_2004_84_115__Valley},
which may completely change the valley physics in QDs.

In this paper, we present a systematic study on the intervalley coupling
strength by QD confinement potentials on extended TMD monolayer. The
numerical calculations used two methods: (i) the envelope function
method (EFM) \cite{Kohn_canonical_transformation,Pantelides_Sah_1974_10_621__Theory,DiVincenzo_Mele_1984_29_1685__Self,Wang_Zunger_1999_59_15806__Linear},
in conjunction with first-principles wavefunctions of the band-edge
Bloch states of the 2D bulk; and (ii) the real-space tight-binding
(RSTB) approach based on a three-band model for monolayer TMDs \cite{Liu_Xiao_2013_88_85433__Three}.
The EFM is limited to circular shaped QDs and is used primarily as
a benchmark for the RSTB approach which can handle arbitrarily shaped
QDs. The intervalley coupling is defined here as the off-diagonal
matrix element, due to the QD potentials, between states with the
same spin but opposite valley index. For circular QD potentials with
various lateral size, potential depth and smoothness, the intervalley
couplings from the two different approaches agree well, which justify
both methods. We then use the RSTB method to address QD potentials
of lower symmetry, including the triangular, hexagonal, and square
shaped QDs, where the intervalley coupling strength is investigated
as functions of lateral size, depth and smoothness of confinement
potentials. 

Our main findings are summarized below. For confinement potentials
with the $C_{3}$ rotational symmetry (like that of the lattice),
both the numerical results and the symmetry analysis show that the
intervalley coupling strength depends sensitively on the central position
of the potential: the coupling is maximized (zero) if the potential
is centered at a M (X) site (cf. figure 1).\textbf{ }The results stated
below refer to the M-centered potentials. The intervalley coupling
as a function of the lateral size of the QDs exhibits fast oscillations
with an nearly exponentially decaying envelop. When the wall of the
potential well changes from a vertical one to a sloping one, the intervalley
coupling strength has a fast decrease by 2-3 orders of magnitude,
and saturates when the length scale of the slope is beyond five lattice
constants. For potentials with sloping walls, the intervalley coupling
increases with increasing the potential depth. For potentials with
vertical walls, the dependence on the potential depth is non-monotonic.
Interestingly, when all parameters are comparable, the intervalley
coupling can be smaller by several orders of magnitude in certain
QD potentials. These include the circular shaped QDs, the triangular
and hexagonal shaped QDs with all sides along the zigzag crystalline
axes. The latter two types of confinement potentials can be relevant
in QDs formed by lateral heterostructures of different TMDs \cite{Huang_Xu_2014____Lateral}.
In contrast to graphene QDs \cite{Trauzettel_Burkard_2007_3_192__Spin}
and silicon QDs \cite{Saraiva_Koiller_2009_80_81305__Physical,Koiller_Das_2004_70_115207__Shallow,Goswami_Eriksson_2007_3_41__Controllable,Friesen_Coppersmith_2010_81_115324__Theory,Culcer_Das_2010_82_155312__Quantum,Friesen_Coppersmith_2007_75_115318__Valley,Nestoklon_Ivchenko_2006_73_235334__Spin,Boykin_Lee_2004_84_115__Valley},
the intervalley coupling here is much smaller as the electron wavefunction
vanishes at the boundary of QDs. For all QDs studied, the largest
intervalley coupling is upper bounded by 0.1 meV, found for small
QDs with diameters of 20 nm and with confinement potentials of vertical
walls. Such intervalley coupling is much smaller compared to the diagonal
energy difference between states with the same spin but opposite valley
index, which arises from the spin-valley coupling and ranges from
several to several tens meV in different monolayer TMDs \cite{Liu_Xiao_2013_88_85433__Three}.
Therefore, our results mean that valley hybridization is in general
negligible, and valley pseudospin in monolayer TMD QDs is a good quantum
degree of freedom as in the 2D bulk, with the optical addressability
for quantum controls. 

The rest of the paper is organized as follows. In section \ref{sec:EFM},
we formulate the EFM for the two-band $\kp$ model of the monolayer
TMDs \cite{Xiao_Yao_2012_108_196802__Coupled}, which can be exactly
solved for circular shaped QDs. We then present the numerical results
for intervalley coupling, calculated in conjunction with the first-principles
wavefunctions. In section \ref{sec:Sym}, we present the symmetry
analysis on the M-centered and X-centered potentials with $C_{3}$
rotational symmetry. Section \ref{sec:RSTB} is on the RSTB approach.
Numerical results by RSTB method for circular QDs are first compared
with the EFM results, and then the RSTB results for various shaped
QDs is presented. The results are qualitatively the same and quantitatively
comparable for different MX$_{2}$, and MoS$_{2}$ is presented as
the example throughout the paper.

\section{Circular QDs in envelope function method\label{sec:EFM}}

\subsection{The model and method}

\begin{figure}
\centering{}\includegraphics[width=15cm]{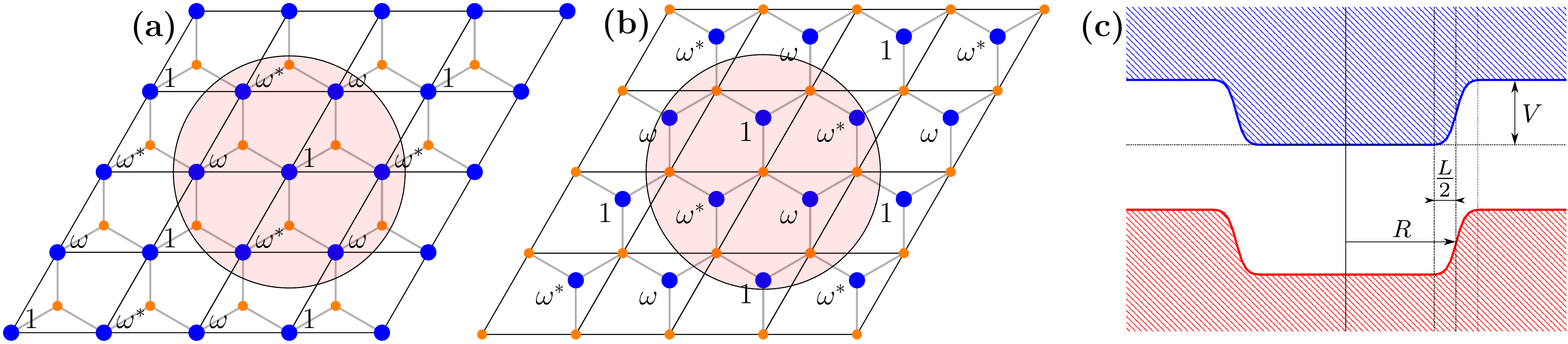}\caption{Schematics of a circular confinement potential on a MoS$_{2}$ monolayer
centered at (a) Mo site (larger blue dot) and (b) S site (smaller
orange dot). $\omega\equiv e^{i\frac{2\pi}{3}}$, $\omega^{*}$ and
$1$ are respectively the projection of the plane wave component of
the Bloch function at $K$ on each Mo site (i.e. $e^{iK\cdot\vr_{{\rm Mo}}}$),
referred as the lattice phase factor in the text. (c) The spatial
profile of the conduction and valence band edges for the circular
potential well with radius $R$ and depth $V$. The length scale $L$
characterizes the smoothness of the potential.\label{fig:well}}
\end{figure}

We consider first a QD defined by a circular shaped potential $U(r)$
in a TMD monolayer. Such confinement potential can be characterized
by three parameters: the lateral size, the depth, and the smoothness
of the potential well. Without loss of generality, we assume the potential
has the form 
\begin{equation}
U(r)=\begin{cases}
0, & r<R-\frac{L}{2}\\
\frac{V}{2}\big[\dfrac{\tanh\frac{C(r-R)}{L/2}}{\tanh C}+1\big], & R-\frac{L}{2}\le r\le R+\frac{L}{2}\\
V, & r>R+\frac{L}{2}
\end{cases}\label{eq:potential}
\end{equation}
where $R$ and $V$ are respectively the lateral size and the depth
of the potential well. $L$ characterizes the smoothness of the potential
well boundary {[}see figure \ref{fig:well}(c){]}. Large $L$ corresponds
to smooth potentials (e.g. in confinement generated by patterned electrodes),
while $L=0$ corresponds to potential well with vertical wall (e.g.
in lateral heterostructures of different TMDs). The parameter $C=2.5$
is used here.

For typical confinement with $R$ much larger than the lattice constants,
the bound states in the QDs are formed predominantly from the band-edge
Bloch states in the $\pm K$ valleys of the 2D bulk. In the EFM, we
start off with the two-band $k\cdot p$ Hamiltonian derived for the
band edges at the $\pm K$ valleys of monolayer TMDs \cite{Xiao_Yao_2012_108_196802__Coupled,Liu_Xiao_2013_88_85433__Three},
where valley is a discrete index. Thus the EFM leads to confined wavefunctions
formed with the $K$ and $-K$ valley Bloch states respectively, denoted
as $\Psi^{\tau,s}$ where $\tau=\pm$ is the valley index denoting
the $\pm K$ valley, and $s=+(\uparrow)$ or $-(\downarrow)$ is the
spin index. Taking these states as the basis, intervalley coupling
here refers to the off-diagonal matrix elements between these states
due to the confinement potential $U(r)$ \cite{Saraiva_Koiller_2009_80_81305__Physical}. 

As confinement potential is spin-independent, intervalley coupling
vanishes between states with opposite spin index. For holes, the band
edges of the 2D bulk are spin-valley locked because of the giant spin-orbit
coupling \cite{zhu_giant_2011,Xiao_Yao_2012_108_196802__Coupled},
i.e. valley $K$ ($-K$) has spin up (down) only. Thus, intervalley
coupling is absent for holes. We concentrate on the intervalley coupling
of confined electron states $\Psi^{+,s}$ and $\Psi^{-,s}$ in the
two valleys with the same spin index, which is defined as 
\begin{equation}
V_{{\rm inter}}^{s}=\bkthree{\Psi^{+,s}}{U(r)}{\Psi^{-,s}}
=
\int\Psi^{+,s}(\vr)^{*}U(r)\Psi^{-,s}(\vr)\,{\rm d}\vr.
\label{eq:Vvcs}
\end{equation}
$V_{\rm inter}^{s}$
  here is the off-diagonal matrix element, due to the QD potentials, between states with the same spin but opposite valley index. This quantity determines to what degree the two bulk valleys can hybridize in the QD confinement, by competing with the diagonal energy difference between $\Psi^{-,s}$
  and $\Psi^{+,s}$. If $V_{\rm inter}^{s}$
  is much smaller than the diagonal energy difference, then the bulk valley index is still a good quantum number in QD. If the coupling is comparable or larger than the diagonal energy difference, then the QD eigenstates will be superpositions of the two bulk valleys. In the limit that $V_{\rm inter}^{s}$
  is much larger than the diagonal energy difference, the QD eigenstates are the symmetric and antisymmetric superpositions of the two valleys, and their energy splitting is given by $2|V_{\rm inter}^s|$.
Due to the time-reversal symmetry, $|V_{{\rm inter}}^{\uparrow}|=|V_{{\rm inter}}^{\downarrow}|$.
Hence we consider here only spin-up case and drop the spin index $s$
in superscripts for brevity. 

To obtain $\Psi^{\tau}$ in EFM, we substitute $-i\nabla$ for $\vk$
in the $k\cdot p$ Hamiltonian of the 2D bulk\cite{Xiao_Yao_2012_108_196802__Coupled,Liu_Xiao_2013_88_85433__Three},
and the confinement potential $U(r)$ is added as an onsite energy
term. The QD Hamiltonian in each valley is then \cite{Hewageegana_Apalkov_2008_77_245426__Electron,Pereira_Peeters_2007_7_946__Tunable,Chen_Chakraborty_2007_98_186803__Fock,Matulis_Peeters_2008_77_115423__Quasibound,Recher_Trauzettel_2009_79_85407__Bound,Pereira_Farias_2009_79_195403__Landau,Hewageegana_Apalkov_2009_79_115418__Trapping}
\begin{eqnarray}
H^{\tau} & = & \begin{bmatrix}\frac{\Delta}{2}+U(r) & at(-i\tau\frac{\partial}{\partial x}-\frac{\partial}{\partial y})\\
at(-i\tau\frac{\partial}{\partial x}+\frac{\partial}{\partial y}) & -\frac{\Delta}{2}+\tau s\lambda+U(r)
\end{bmatrix},\label{eq:Htau}
\end{eqnarray}
where $\Delta\sim2$ eV is the bulk bandgap of TMD monolayer, $a$
is the lattice constant, and $t$ is the effective hopping. The bases
of the Hamiltonian  \eqref{eq:Htau} are the conduction and valence
Bloch states at $\pm K$ point, $\varphi_{\alpha}^{\tau}(\vr)=e^{i\tau\vK\cdot\vr}u_{\alpha}^{\tau}(\vr)$
($\alpha=$c, v and $u_{\alpha}^{\tau}$ is the cell periodic part),
and the eigenfunction of \eqref{eq:Htau} is a spinor of envelope
function
\begin{equation}
\psi^{\tau}=\begin{bmatrix}\psi_{{\rm c}}^{\tau}(\vr)\\
\psi_{{\rm v}}^{\tau}(\vr)
\end{bmatrix},
\end{equation}
which satisfy the Schrodinger equation $H^{\tau}\psi^{\tau}=E\psi^{\tau}$.
The overall wavefunctions of the QD states are then
\begin{equation}
\Psi^{\tau}(\vr)=\psi_{{\rm c}}^{\tau}(\vr)\varphi_{{\rm c}}^{\tau}(\vr)+\psi_{{\rm v}}^{\tau}(\vr)\varphi_{{\rm v}}^{\tau}(\vr).
\end{equation}

Since $U(r)$ has the rotational symmetry, it is easy to verify that
the eigen spinor $\psi^{\tau}$ is of the form 
\begin{equation}
\psi^{\tau}=\begin{bmatrix}b_{{\rm c}}^{\tau}(r)e^{im\theta}\\
ib_{{\rm v}}^{\tau}(r)e^{i(m+\tau)\theta}
\end{bmatrix},\label{eq:psib}
\end{equation}
where $m$ is integer, and $(r,\theta)$ are the polar coordinates
of $\vr$. $b_{{\rm c}}^{\tau}$ and $b_{{\rm v}}^{\tau}$ are real
radial functions that satisfy

\begin{equation}
\left\{ \begin{array}{lc}
{\displaystyle \tau\frac{\partial b_{{\rm c}}^{\tau}}{\partial r}-\frac{m}{r}b_{{\rm c}}^{\tau}=[-\frac{\Delta}{2}+s\tau\lambda+U(r)-E]b_{{\rm v}}^{\tau}} & \vphantom{\Bigg[}\\
{\displaystyle \tau\frac{\partial b_{{\rm v}}^{\tau}}{\partial r}+\frac{m+\tau}{r}b_{{\rm v}}^{\tau}=-[\frac{\Delta}{2}+U(r)-E]b_{{\rm c}}^{\tau}}
\end{array}\right.\label{eq:Eqbcbv}
\end{equation}
$b_{{\rm c}}^{\tau}$ and $b_{{\rm v}}^{\tau}$ are Bessel functions
in the region where $U(r)$ is a constant ($r<R-\frac{L}{2}$ or $r>R+\frac{L}{2}$),
and need to be solved numerically in the region $R-\frac{L}{2}\le r\le R+\frac{L}{2}$.
Boundary conditions are then matched at $r=R-\frac{L}{2}$ and $r=R+\frac{L}{2}$
to determine the solutions and eigenvalues. Bound states are found
to exist in the energy range $(\frac{\Delta}{2},\frac{\Delta}{2}+V)$,
as expected. Here $t$ is used as the unit of energy and $a$ as the
unit of length. 

\begin{figure}
\begin{centering}
\includegraphics[width=14cm]{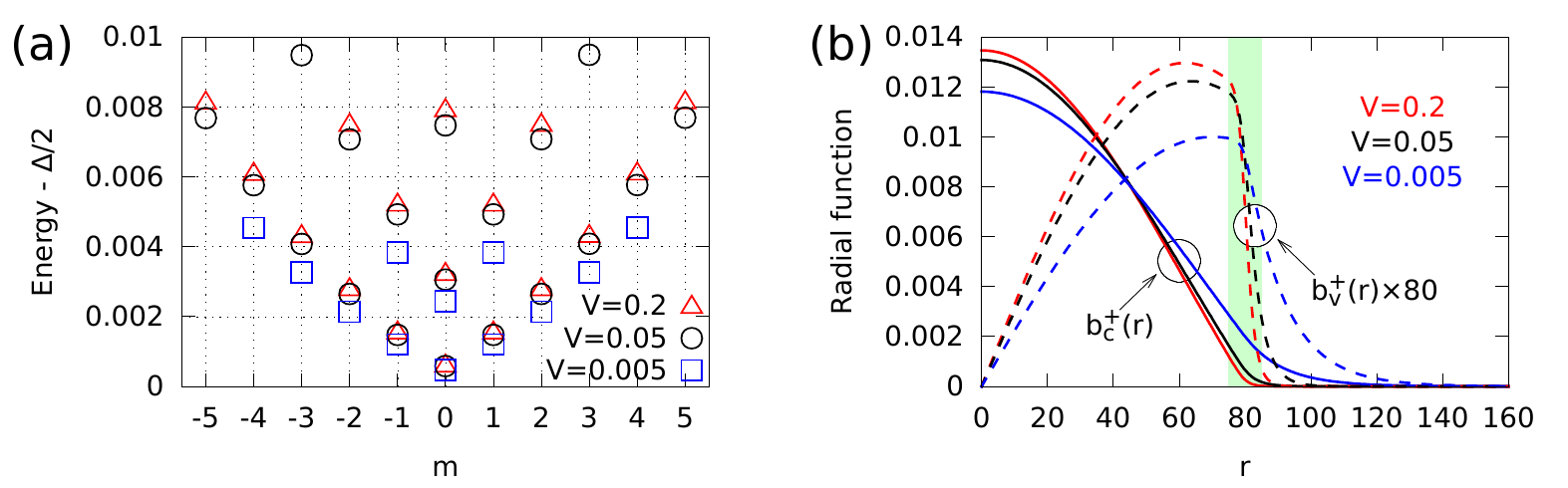}
\par\end{centering}

\caption{Energy levels and envelope functions for circular QDs calculated with
EFM. (a) Energy levels of the lowest several bound states measured
from the bulk band edge. Three different potentials with $V=$0.2,
0.05, and 0.005 respectively are shown for comparison. Other parameters
are $R=80$ and $L=10$. (b) The radial envelope functions $b_{{\rm c}}^{+}(r)$
and $b_{{\rm v}}^{+}(r)$ {[}see equation \eqref{eq:psib}{]} for
the ground states in (a), i.e. the lowest-energy ones with $m=0$.
Note that $b_{{\rm v}}^{+}(r)$ are magnified by a factor of 80 for
clarity. The green shaded region corresponds to the slopping wall
of the potential well. Units for energy and length are $t$ and $a$
respectively.\label{fig:ene}}
\end{figure}

The intervalley coupling defined in Eq. (\ref{eq:Vvcs}) can then be expressed as
\begin{eqnarray}
V_{{\rm inter}} & = & \bkthree{\sum_{\alpha}\psi_{\alpha}^{+}e^{i\bm{K}\cdot\vr}u_{\alpha}^{+}}U{\sum_{\beta}\psi_{\beta}^{-}e^{-i\vK\cdot\vr}u_{\beta}^{-}}=\sum_{\alpha,\beta}V_{\alpha\beta}\ ,\label{eq:Vinter}
\end{eqnarray}
where

\begin{equation}
V_{\alpha\beta}=\sum_{\Delta\bm{G}}S_{\alpha\beta}(\Delta\bm{G})I_{\alpha\beta}(\Delta\bm{G}),\label{eq:Vab}
\end{equation}

\begin{equation}
I_{\alpha\beta}(\Delta\bm{G})=\int\psi_{\alpha}^{+}(\vr)^{*}\psi_{\beta}^{-}(\vr)e^{i(\Delta\bm{G}-2\bm{K})\cdot\vr}U(r)\,{\rm d}\vr,\label{eq:IdG}
\end{equation}
and
\begin{eqnarray}
S_{\alpha\beta}(\Delta\bm{G}) & = & \sum_{\bm{G}}c_{\alpha}^{+}(\bm{G})^{*}c_{\beta}^{-}(\bm{G}+\Delta\bm{G}).\label{eq:SdG}
\end{eqnarray}
In the above equations, $\alpha,\beta\in$\{c, v\}, and $c_{\alpha}^{\tau}(\bm{G})$
are the Fourier coefficients of the periodic part of the Bloch function
$u_{\alpha}^{\tau}(\vr)=\sum_{\bm{G}}c_{\alpha}^{\tau}(\bm{G})e^{i\bm{G}\cdot\vr}$.
$\bm{G}$ is a reciprocal lattice vector (RLV), and $\Delta\bm{G}=\bm{G}'-\bm{G}$
is also a RLV. Here we are interested in the QD ground states only,
which correspond to the $m=0$ case in equation \eqref{eq:IdG} {[}see
figure \ref{fig:ene}(a){]}. The coefficients $c_{\alpha}^{\tau}(\bm{G})$
are obtained here from the first-principles all-electron wavefunctions
calculated by the ABINIT package \cite{abinit1,note2}.

We note that the Hamiltonian in equation \eqref{eq:Htau} does not
contain the conduction band spin-valley interaction \cite{Liu_Xiao_2013_88_85433__Three,Kormanyos_Falko_2013_88_45416__Monolayer},
which is of the Ising form that changes only the diagonal energies
but not the wavefunctions of the states $\Psi^{+,s}$ and $\Psi^{-,s}$.
Thus, this spin-valley interaction does not affect the intervalley
coupling strength defined here, but it leads to an energy difference
between states $\Psi^{+,s}$ and $\Psi^{-,s}$ that ranges from several
to several tens meV \cite{Liu_Xiao_2013_88_85433__Three,Kormanyos_Falko_2013_88_45416__Monolayer}.
Valley hybridization in the QDs is determined by the competition between
$|V_{{\rm inter}}|$ and this diagonal energy difference.

\subsection{Numerical results \label{sub:Numerical-results}}

In this subsection we present the numerical calculation results. The
parameters from   \cite{Xiao_Yao_2012_108_196802__Coupled} are used:
$t=1.1\,$eV, $a=3.193\,$\AA{}, $\Delta=1.66\,$eV, and $2\lambda=0.15\,$eV.
We consider potentials centered at a Mo site, as shown in figure \ref{fig:well}(a).
As we will prove in Section \ref{sec:Sym}, the intervalley coupling
is strongest for such Mo-centered confinement potential, and will
vanish if the circular confinement potential is centered at a S site.

\subsubsection{Energy levels and bound states}

We first look at the energy eigenvalues and eigenstates of equation
\eqref{eq:Htau}, which give the energy spectrum in the circular QD
confinement, and the radial envelope functions to calculate the intervalley
coupling {[}cf. equation \eqref{eq:Vvcs}{]}. The energy spectra are
shown in figure \ref{fig:ene}(a) for $R=80$, $L=10$ and three well
depths$V=$0.2, 0.05, and 0.005. For $V=0.2$ and 0.05, only the low-energy
parts of the spectra are shown. $m$ is the azimuthal quantum number
of the envelope function defined in equation \eqref{eq:psib}. Obviously,
the ground state has $m=0$. The spectra shown are for the $K$ valley
($\tau=1$) and it is obvious from equation \eqref{eq:Eqbcbv} that
states of $\tau$, $s$, $m$ are degenerate with states of $-\tau,$
$-s$, $-m$. For the ground states, the radial envelope functions
$b_{{\rm c}}^{+}(r)$ and $b_{{\rm v}}^{+}(r)$ are shown in figure
\ref{fig:ene}(b), from which we can see that they are smooth and
localized within the well and the $b_{{\rm v}}^{+}(r)$ components
are orders of magnitude smaller compared to $b_{{\rm c}}^{+}(r)$.
This means that for electron confined in the quantum dot, the wavefunction
is predominantly contributed by the conduction band edge Bloch functions,
and the contribution from the valence band edge is negligible. In
general, the Hamiltonian  \eqref{eq:Htau} shall be solved for addressing
confinement of Dirac fermions, as we did here. Nevertheless, as shown
here, the much simplified approach of effective mass approximation
can be well justified for monolayer TMDs with the large bandgap. Importantly,
the wavefunction has vanishing amplitude at the QD boundary {[}green
shaded region in figure \ref{fig:ene}(b){]}. This is in contrast
to the wavefunctions in graphene QDs \cite{Trauzettel_Burkard_2007_3_192__Spin}
and silicon QDs \cite{Saraiva_Koiller_2009_80_81305__Physical,Koiller_Das_2004_70_115207__Shallow,Goswami_Eriksson_2007_3_41__Controllable,Friesen_Coppersmith_2010_81_115324__Theory,Culcer_Das_2010_82_155312__Quantum,Friesen_Coppersmith_2007_75_115318__Valley,Nestoklon_Ivchenko_2006_73_235334__Spin,Boykin_Lee_2004_84_115__Valley}
where significant valley hybridization is found.

\subsubsection{The intervalley coupling}

\begin{figure}
\centering{}\includegraphics[width=15cm]{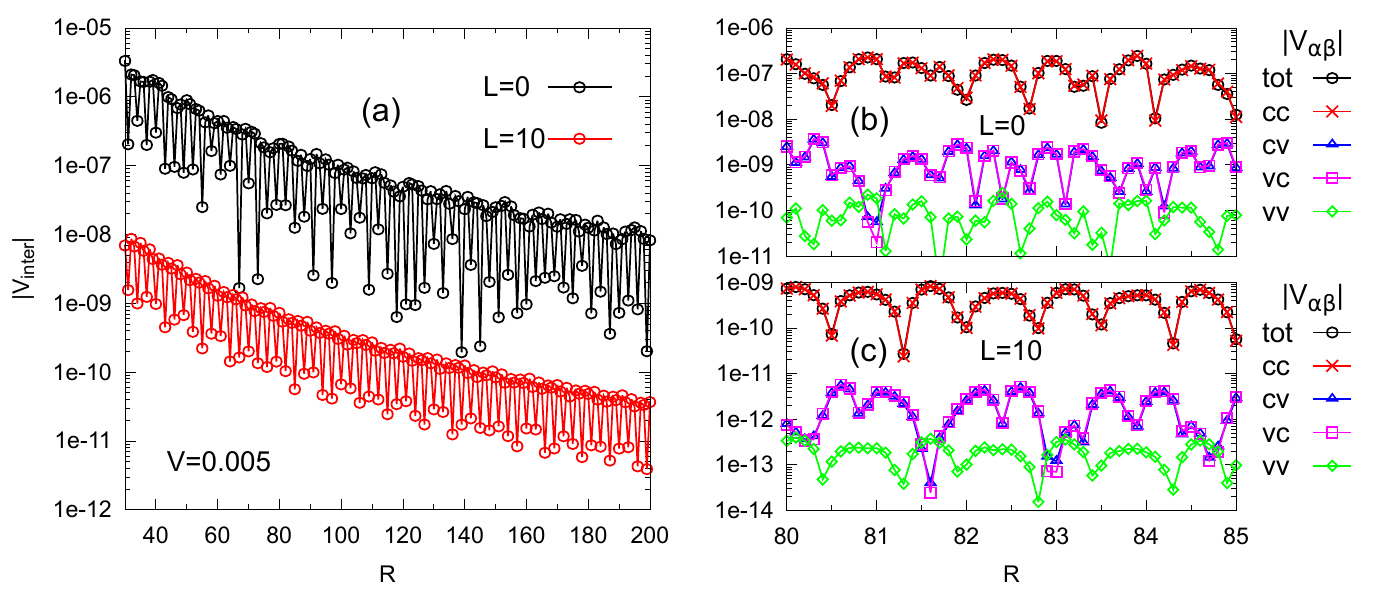}\caption{(a) The magnitude of the intervalley coupling $|V_{{\rm inter}}|$
vs the radius $R$ for circular QDs with $L=0$ (black) and $L=10$
(red), and $V=0.005$ calculated in EFM. (b) and (c) are the zoom
in for the parameter range $R\in[80,85]$. Black circle denotes the
overall $|V_{{\rm inter}}|$. Other symbols denote the contributions
$|V_{\alpha\beta}|$ {[}$\alpha\beta=$cc (red cross), cv (blue triangle),
vc (magenta square), and vv (green diamond){]} to the intervalley
coupling, for $L=0$ and 10 respectively. See equations \eqref{eq:Vinter},
\eqref{eq:Vab}, and \eqref{eq:IdG} in the text for definitions.
All lengths are in the unit of $a$ and energies in the unit of $t$.\label{fig:Vvc-R}}
\end{figure}

\begin{figure}
\centering{}\includegraphics[width=8cm]{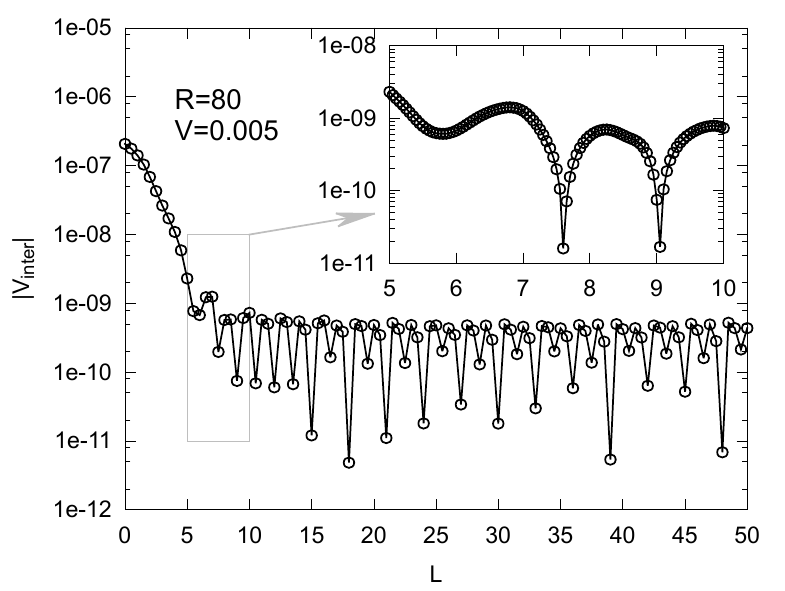}\caption{The magnitude of the intervalley coupling $|V_{{\rm inter}}|$ vs
$L$ for circular QDs with $R=80$, $V=0.005$ calculated in EFM.
Inset shows the details of the main figure in the range of $L\in[5,10]$.
All lengths are in the unit of $a$ and energies in the unit of $t$.\label{fig:Vvc-L}}
\end{figure}

With the envelope functions given above, and the band-edge Bloch functions
from the ABINIT package \cite{abinit1,note2}, we can calculate the
numerical values for the intervalley coupling {[}cf. equation \eqref{eq:Vinter}{]}, 

Figure \ref{fig:Vvc-R} shows the intervalley coupling strength $|V_{{\rm inter}}|$
versus the radius of the potential $R$ for circular QDs with fixed
smoothness ($L=0$ and 10) and depth ($V=0.005$) of potential {[}cf.
equation \eqref{eq:potential}{]}. The curve with $L=10$ is more
than two orders of magnitude smaller than the one with $L=0$, which
can also be seen in figure \ref{fig:Vvc-L} and discussed in detail
later. From figure \ref{fig:Vvc-R}(b) and (c), we can see that the
dominant contribution to the intervalley coupling is from $|V_{{\rm cc}}|$,
which is orders of magnitude larger than $|V_{{\rm vv}}|$, $|V_{{\rm cv}}|$,
$|V_{{\rm vc}}|$ {[}cf. equation \eqref{eq:Vab}{]}, consistent with
the fact that the wavefunction of confined electron is predominantly
from the conduction band edge Bloch functions. Figures \ref{fig:Vvc-R}(b)
and (c) also clearly show that $|V_{{\rm inter}}|$ oscillates vs
$R$ in the scale of the lattice constant $a$. The oscillation with
the size of QD is a generic feature of intervalley coupling, due to
the large momentum space separation of the two valleys, which have
been noted in previous works \cite{Boykin_Lee_2004_84_115__Valley,Nestoklon_Ivchenko_2006_73_235334__Spin,Friesen_Coppersmith_2007_75_115318__Valley}.
Apart from these fast oscillations, the intervalley coupling strength
decreases with the increase of $R$ in a nearly exponential way.

\begin{figure}
\centering{}\includegraphics[width=10cm]{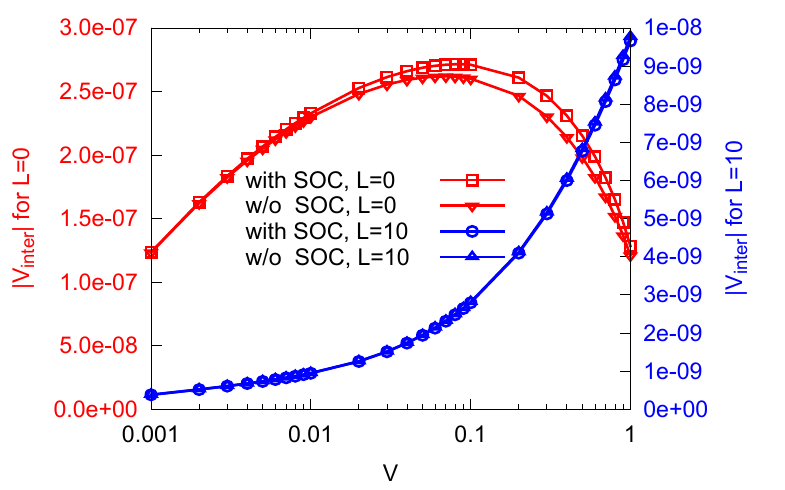}\caption{Intervalley coupling strength $|V_{{\rm inter}}|$ as a function of
the potential well depth $V$ for circular QD with $R=80$, $L=0$
(red $\square$ and $\triangledown$, left axis) and $L=10$ (blue
$\circ$ and $\vartriangle$, right axis) calculated in EFM. Results
with (red $\square$, blue $\circ$ ) and without SOC (red $\triangledown$,
blue $\vartriangle$) are compared. All lengths are in the unit of
$a$ and energies in the unit of $t$. \label{fig:Vvc-V}}
\end{figure}

Figure \ref{fig:Vvc-L} shows the $|V_{{\rm inter}}|$ as a function
of $L$, the smoothness of the potential, for circular QDs with $R=80$
and $V=0.005$. Largest intervalley coupling is found at $L=0$ that
corresponds to a circular square well, where the potential has a sudden
jump at the QD boundary. It can be seen that $|V_{{\rm inter}}|$
decreases monotonically as $L$ increases up to a value of 5, and
then it starts to oscillate rapidly around a constant value. The inset
is the zoom in for $5\le L\le10$, which shows that the oscillation
is also in the length scale of lattice constant, similar to the $|V_{{\rm inter}}|$
vs. $R$ relation. 

Figure \ref{fig:Vvc-V} shows the dependence of $|V_{{\rm inter}}|$
on the potential depth $V$, when $R=80$ and $L=0$ and 10. We can
see that $|V_{{\rm inter}}|$ increases with increasing $V$ when
$L=10$, and it first increases and then decreases with increasing
$V$ when $L=0$. In this figure, we also compare the intervalley
couplings calculated with and without spin-orbit coupling (SOC). For
the calculation without SOC, we set $s=0$ in equation \eqref{eq:Htau},
and the plane-wave coefficients in equation \eqref{eq:SdG} are also
calculated without including the SOC in the ABINIT package. In figure
\ref{fig:Vvc-V}, it is clear that intervalley couplings with and
without SOC coincide when $L=10$, and have small difference when
$L=0$. We also compared the $|V_{{\rm inter}}|$ vs. $R$ curve (fixing
$L=10$, $V=0.005$) and $|V_{{\rm inter}}|$ vs. $L$ curve (fixing
$R=80$, $V=0.005$), for the calculations with and without SOC, and
the differences are also negligible. This means that SOC has negligible
effects on $|V_{{\rm inter}}|$. This is well expected as $|V_{{\rm inter}}|$
here is just the off-diagonal matrix element of the confinement potential,
between the QD wavefunctions in the $K$ and $-K$ valleys. The SOC
in the $K$ valleys is of the longitudinal form \cite{Xiao_Yao_2012_108_196802__Coupled},
which does not affect the wavefunction. The SOC, however, will cause
spin splittings and hence affect the spectral of quantum dot states
in the $K$ and $-K$ valleys {[}cf. figure \ref{fig:ene}(a){]}.
When we consider the effect of intervalley coupling on the quantum
dot states, the off-diagonal matrix element $|V_{{\rm inter}}|$ will
then compete with the differences in these diagonal energies, and
the SOC induced spin splitting will play an important role.

From the dependence of $|V_{{\rm inter}}|$ on $R$, $L$, and $V$
discussed above, we find that the intervalley coupling strength in
a circular confinement potential varies over several orders of magnitude,
depending on the length scale, smoothness, and depth of the confinement.
The magnitude of the intervalley coupling is small on the whole, in
the order of $\mu$eV or below. This is much smaller compared to intervalley
coupling in QDs formed in silicon inversion layers \cite{Saraiva_Koiller_2009_80_81305__Physical,Nestoklon_Ivchenko_2006_73_235334__Spin,Koiller_Das_2004_70_115207__Shallow,Goswami_Eriksson_2007_3_41__Controllable,Friesen_Coppersmith_2007_75_115318__Valley,Friesen_Coppersmith_2010_81_115324__Theory,Culcer_Das_2010_82_155312__Quantum,Boykin_Lee_2004_84_115__Valley}.

\section{Symmetry analysis for QDs with $C_{3}$ rotational symmetry\label{sec:Sym}}

In the above section, we have seen that the intervalley coupling is
very sensitive to the geometry of the confinement potential, as evident
from the fast oscillations of $|V_{{\rm inter}}|$ versus $R$ or
$L$. Here, based on symmetry analysis, we will show that the intervalley
coupling strength $|V_{{\rm inter}}|$ also depends on the position
of the potential center, in a way that $V_{{\rm inter}}$ is largest
when the potential is centered at Mo atom {[}figure \ref{fig:well}(a){]}
but vanishes if centered at S atom {[}figure \ref{fig:well}(b){]}.
To explain this, we analyze $V_{\alpha\beta}$ {[}see equation \eqref{eq:Vab}{]}
by examining the behavior of $S_{\alpha\beta}(\Delta\bm{G})$ and
$I_{\alpha\beta}(\Delta\bm{G})$ under the $C_{3}$ rotation. 

Using the Fourier relation $c_{\alpha}^{\tau}(\bm{G})=\frac{1}{\Omega}\int_{\Omega}u_{\alpha}^{\tau}(\vr)e^{-i\bm{G}\cdot\vr}\,{\rm d}\vr$
in which $\Omega$ is the area of the 2D cell, $S_{\alpha\beta}(\Delta\bm{G})$
can be rewritten as 
\begin{eqnarray}
S_{\alpha\beta}(\Delta\bm{G}) & = & \frac{1}{\Omega}\int_{\Omega}u_{\alpha}^{+}(\vr)^{*}u_{\beta}^{-}(\vr)e^{-i\Delta\bm{G}\cdot\vr}\,{\rm d}\vr\label{eq:SdGa}\\
 & = & \frac{1}{\Omega}\int_{\Omega}\varphi_{\alpha}^{+}(\vr)^{*}\varphi_{\beta}^{-}(\vr)e^{-i\bm{g}\cdot\vr}\,{\rm d}\vr,\ \ \ \ \ \ \label{eq:SdGb}
\end{eqnarray}
where $\bm{g}\equiv\Delta\bm{G}-2\vK$. Equation \eqref{eq:SdGa}
shows that its integrand is a cell periodic function. This periodicity
leads to the invariance of $S_{\alpha\beta}$ under the $C_{3}$ rotation
of its integrand. For convenience, we rewrite $\mathbb{S}_{\alpha\beta}(\bm{g})\equiv S_{\alpha\beta}(\Delta\bm{G})$.
Using the form of integrand in equation \eqref{eq:SdGb}, we get 
\begin{eqnarray}
\mathbb{S}_{\alpha\beta}(\bm{g}) & = & \frac{1}{\Omega}\int_{\Omega}C_{3}\Big[\varphi_{\alpha}^{+}(\vr)^{*}\varphi_{\beta}^{-}(\vr)e^{-i\bm{g}\cdot\vr}\Big]\,{\rm d}\vr\nonumber \\
 & = & \frac{1}{\Omega}\int_{\Omega}\big[C_{3}(\varphi_{\alpha}^{+})^{*}\big]\big[C_{3}\varphi_{\beta}^{-}\big]e^{-i\bm{g}\cdot C_{3}^{-1}\vr}\,{\rm d}\vr\nonumber \\
 & = & (\gamma_{\alpha}^{+})^{*}\gamma_{\beta}^{-}\frac{1}{\Omega}\int_{\Omega}(\varphi_{\alpha}^{+})^{*}\varphi_{\beta}^{-}e^{-iC_{3}\bm{g}\cdot\vr}\,{\rm d}\vr\nonumber \\
 & = & (\gamma_{\alpha}^{+})^{*}\gamma_{\beta}^{-}\mathbb{S}_{\alpha\beta}(C_{3}\bm{g}),\label{eq:SC3g}
\end{eqnarray}
where $\gamma_{\alpha}^{\tau}\equiv[C_{3}\varphi_{\alpha}^{\tau}(\vr)]/\varphi_{\alpha}^{\tau}(\vr)$
is the eigenvalue of $C_{3}$ operation corresponding to the Bloch
function $\varphi_{\alpha}^{\tau}(\vr)$. 

For the integral $I_{\alpha\beta}$, we also rewrite it as $\mathbb{I}_{\alpha\beta}(\bm{g})\equiv I_{\alpha\beta}(\Delta\bm{G}-2\vK)$.
Using the integral form of the Bessel function of order $n$ (integer)
$J_{n}(x)=\frac{1}{2\pi}\int_{-\pi}^{\pi}e^{i(x\sin\phi-n\phi)}\,{\rm d}\phi$,
we can simplify $\mathbb{I}_{\alpha\beta}(\bm{g})$ to single integrals
as
\begin{equation}
\mathbb{I}_{{\rm cc}}(\bm{g})=2\pi\int_{0}^{\infty}(b_{{\rm c}}^{+})^{*}U(r)b_{{\rm c}}^{-}J_{0}(|\bm{g}|r)r\,{\rm d}r,\label{eq:Icc}
\end{equation}
\begin{equation}
\mathbb{I}_{{\rm cv}}(\bm{g})=-2\pi e^{-i\phi}\int_{0}^{\infty}(b_{{\rm c}}^{+})^{*}U(r)b_{{\rm v}}^{-}J_{1}(|\bm{g}|r)r\,{\rm d}r,
\end{equation}
\begin{equation}
\mathbb{I}_{{\rm vc}}(\bm{g})=2\pi e^{-i\phi}\int_{0}^{\infty}(b_{{\rm v}}^{+})^{*}U(r)b_{{\rm c}}^{-}J_{1}(|\bm{g}|r)r\,{\rm d}r,
\end{equation}
\begin{equation}
\mathbb{I}_{{\rm vv}}(\bm{g})=-2\pi e^{-i2\phi}\int_{0}^{\infty}(b_{{\rm v}}^{+})U(r)b_{{\rm v}}^{-}J_{2}(|\bm{g}|r)r\,{\rm d}r,\label{eq:Ivv}
\end{equation}
where $\phi$ is the polar angle of the 2D vector $\bm{g}$. It is
obvious that under $\bm{g}\rightarrow C_{3}\bm{g}$, the integral
parts of equations \eqref{eq:Icc}--\eqref{eq:Ivv} do not change,
and hence the ratio $\eta_{\alpha\beta}\equiv\mathbb{I}_{\alpha\beta}(C_{3}\bm{g})/\mathbb{I}_{\alpha\beta}(\bm{g})$
is determined only by the prefactor in front of the integrals:
\begin{equation}
\eta_{{\rm cc}}=1,\ \ \ \ \eta_{{\rm cv}}=\eta_{{\rm vc}}=e^{-i\frac{2\pi}{3}},\ \ \ \ \eta_{{\rm vv}}=e^{i\frac{2\pi}{3}}.\label{eq:etaxx}
\end{equation}
Because the set of $\bm{g}$'s, $\mathbb{G}$, also has the $C_{3}$
symmetry, we can choose one third elements of $\mathbb{G}$, denoted
as $\mathbb{G}_{\frac{1}{3}}$, such that $\mathbb{G}=\sum_{n=0}^{2}C_{3}^{n}\mathbb{G}_{\frac{1}{3}}$.
Using equations \eqref{eq:SC3g} and \eqref{eq:etaxx}, we have 
\begin{eqnarray}
V_{\alpha\beta} & = & \sum_{\bm{g}\in\mathbb{G}_{\frac{1}{3}}}\sum_{n=0}^{2}\Big[\mathbb{S}_{\alpha\beta}(C_{3}^{n}\bm{g})\mathbb{I}_{\alpha\beta}(C_{3}^{n}\bm{g})\Big]\nonumber \\
 & = & \sum_{\bm{g}\in\mathbb{G}_{\frac{1}{3}}}\sum_{n=0}^{2}\Big\{[(\gamma_{\alpha}^{+})^{*}\gamma_{\beta}^{-}]^{-n}\mathbb{S}_{\alpha\beta}(\bm{g})\eta_{\alpha\beta}^{n}\mathbb{I}_{\alpha\beta}(\bm{g})\Big\}\nonumber \\
 & = & (1+\lambda_{\alpha\beta}+\lambda_{\alpha\beta}^{2})\sum_{\bm{g}\in\mathbb{G}_{\frac{1}{3}}}\mathbb{S}_{\alpha\beta}(\bm{g})\mathbb{I}_{\alpha\beta}(\bm{g}),\label{eq:Vxx}
\end{eqnarray}
where 
\begin{equation}
\lambda_{\alpha\beta}=[(\gamma_{\alpha}^{+})^{*}\gamma_{\beta}^{-}]^{-1}\eta_{\alpha\beta}.\label{eq:lmdxx}
\end{equation}
$\eta_{\alpha\beta}$ is already given in equation \eqref{eq:etaxx}
and the rest is to determine $\gamma_{\alpha}^{\tau}$. 

For MoS$_{2}$ monolayer, the conduction and valence band edge Bloch
functions at $\pm K$ points are formed predominantly by the $d_{0}$
(i.e. $d_{z^{2}}$) and $d_{\pm2}$ {[}i.e. $\frac{1}{\sqrt{2}}(d_{x^{2}-y^{2}}\pm id_{xy})${]}
orbitals of Mo atoms respectively \cite{Xiao_Yao_2012_108_196802__Coupled,Liu_Xiao_2013_88_85433__Three,note2}.
There are two contributions to the eigenvalue $\gamma_{\alpha}^{\tau}$
of the $C_{3}$ rotation on the Bloch function $\varphi_{\alpha}^{\tau}(\vr)=e^{i\tau\vK\cdot\vr}u_{\alpha}^{\tau}(\vr)$
(details in Appendix \ref{sec:C3rot}): (i) the rotation of atomic
orbital around its own center, i.e. $C_{3}d_{0}(\vr)=d_{0}(\vr)$
and $C_{3}d_{\pm2}(\vr)=e^{\pm i\frac{2\pi}{3}}d_{\pm2}(\vr)$; (ii)
the change of lattice phase, defined as the value of the planewave
phase factor $e^{i\tau\vK\cdot\vr}$ at each lattice site (cf. figure
\ref{fig:well} for the case of $\tau=+1$). The change of the lattice
phase factor under $C_{3}$ rotation depends on the rotation center.
If the rotation center is at Mo atom as shown in figure \ref{fig:well}(a),
the $C_{3}$ rotation does not change the lattice phase factors leaving
only contribution (i) taking effect, and we obtain 
\begin{equation}
\gamma_{{\rm c}}^{\tau}=1\ \ \ \text{and}\ \ \ \gamma_{{\rm v}}^{\tau}=e^{i\tau\frac{2\pi}{3}}\ \ \ \ (\text{for Mo center}).\label{eq:C3Mo}
\end{equation}
However, if the rotation center is at S atom, as shown in figure \ref{fig:well}(b),
both contributions (i) and (ii) take effect and we have
\begin{equation}
\gamma_{{\rm c}}^{\tau}=e^{i\tau\frac{2\pi}{3}}\ \ \ \text{and}\ \ \ \gamma_{{\rm v}}^{\tau}=e^{-i\tau\frac{2\pi}{3}}\ \ \ \ (\text{for S center}).\label{eq:C3S}
\end{equation}
Putting equations \eqref{eq:etaxx} and \eqref{eq:C3Mo} {[}or \eqref{eq:C3S}{]}
into equation \eqref{eq:lmdxx}, we obtain that $\lambda_{\alpha\beta}=1$
for Mo centered potential and $\lambda_{\alpha\beta}=e^{-i\frac{2\pi}{3}}$
for S centered potential, and the final intervalley coupling is 
\begin{equation}
V_{{\rm inter}}=\begin{cases}
{\displaystyle 3\sum_{\alpha\beta}\sum_{\bm{g}\in\mathbb{G}_{\frac{1}{3}}}\mathbb{S}_{\alpha\beta}(\bm{g})\mathbb{I}_{\alpha\beta}(\bm{g})}, & \text{for Mo centered potential}\\
0, & \text{for S centered potential}
\end{cases}.\label{eq:VvcMoS}
\end{equation}

From the above symmetry analysis, we conclude that the intervalley
coupling induced by a circular confinement potential is sensitive
to the center of the potential. The intervalley coupling is enhanced
when the potential is centered at Mo atom, while it vanishes if the
potential is centered at S atom. This is due to the dependence of
$\gamma_{{\rm \alpha}}^{\tau}$, the eigenvalue of the Bloch function
$\varphi_{\alpha}^{\tau}$ under the $C_{3}$ rotation, on the location
of rotation center. Below, we show that the same conclusions can be
drawn as long as the confinement potential has the $C_{3}$ symmetry,
i.e. intervalley coupling is strongest (zero) if the center of the
potential is at Mo (S) site.

\subsection{Noncircular QD with $C_{3}$ symmetry}

Consider a noncircular QD confinement potential with $C_{3}$ symmetry
only, $C_{3}U(\vr)=U(\vr)$. In such potential that lacks the circular
symmetry, we do not have exact solution in general for the massive
Dirac Fermion Hamiltonian in equation \eqref{eq:Htau}. Nevertheless,
the numerical results presented in section \ref{sub:Numerical-results}
have well justified the effective mass approximation, where we can
construct the QD electron wavefunction based on the conduction band
edge Bloch function only. The ground-state envelope function $\psi_{{\rm c}}^{\tau}$
can be obtained by solving the following effective-mass Hamiltonian
\begin{equation}
H_{{\rm em}}=-\frac{\hbar^{2}}{2m_{{\rm eff}}}\nabla^{2}+U(\vr),
\end{equation}
where $m_{{\rm eff}}$ is the effective mass of conduction band at
$\pm K$. With the same $m_{{\rm eff}}$ at $\pm K$, the envelope
functions in the two valleys are the same: $\psi_{{\rm c}}^{+}=\psi_{{\rm c}}^{-}$.
To analyze the symmetry, we need to know $\gamma_{{\rm c}}^{\tau}$
and $\eta_{{\rm cc}}$ {[}see equations \eqref{eq:Vxx} and \eqref{eq:lmdxx}{]}.
$\gamma_{{\rm c}}^{\tau}$ is independent of the shape of the QD,
and equations \eqref{eq:C3Mo} and \eqref{eq:C3S} are still valid
here. $\eta_{{\rm cc}}$ is determined by the symmetry of the envelope
function $\psi_{{\rm c}}^{\tau}$. Since $U(\vr)$ is symmetric under
$C_{3}$ rotation, the non-degenerate ground-state envelope function
$\psi_{{\rm c}}^{\tau}$ in each valley has to be an eigenstate of
$C_{3}$, i.e. $C_{3}\psi_{{\rm c}}^{\tau}=\rho\psi_{{\rm c}}^{\tau}$
where $|\rho|=1$. Using equation \eqref{eq:IdG} and $\rho^{*}\rho=1$,
we have
\begin{eqnarray}
\mathbb{I}_{{\rm cc}}(C_{3}\bm{g}) & = & \int\psi_{{\rm c}}^{+}(\vr)^{*}\psi_{{\rm c}}^{-}(\vr)e^{iC_{3}\bm{g}\cdot\vr}U(\vr)\,{\rm d}\vr=\int[\rho\psi_{{\rm c}}^{+}(\vr)]^{*}[\rho\psi_{{\rm c}}^{-}(\vr)]e^{i\bm{g}\cdot C_{3}^{-1}\vr}U(\vr)\,{\rm d}\vr\nonumber \\
 & = & \int[C_{3}\psi_{{\rm c}}^{+}(\vr)]^{*}[C_{3}\psi_{{\rm c}}^{-}(\vr)]e^{i\bm{g}\cdot C_{3}^{-1}\vr}[C_{3}U(\vr)]\,{\rm d}\vr=\int C_{3}[\psi_{{\rm c}}^{+}(\vr)^{*}\psi_{{\rm c}}^{-}(\vr)e^{i\bm{g}\cdot\vr}U(\vr)]\,{\rm d}\vr\nonumber \\
 & = & \int\psi_{{\rm c}}^{+}(\vr)^{*}\psi_{{\rm c}}^{-}(\vr)e^{i\bm{g}\cdot\vr}U(\vr)\,{\rm d}\vr=\mathbb{I}_{{\rm cc}}(\bm{g})
\end{eqnarray}
Thus $\eta_{{\rm cc}}=1$, consistent with equation \eqref{eq:etaxx}.
The conclusion of equation \eqref{eq:VvcMoS} is therefore valid for
QD confinement potential with $C_{3}$ symmetry.

In summary, the band edge Bloch functions at $K$ and $-K$ have the $C_{3}$ 
  rotational symmetry of the lattice. However, they transform differently when the rotation center is Mo site or S site, as shown in Figure 1a and 1b respectively. When the QD confinement potential also has the $C_{3}$
  rotational symmetry about a S site, the rotational symmetry of the band edge Bloch functions about the S site leads to destructive interference and hence vanishing intervalley coupling. This is the physical origin of the dependence of intervalley coupling on the central position of QD potential.

\section{Real-space tight-binding method\label{sec:RSTB}}

In the EFM, except for the circular shaped confinement potential,
it is difficult to find the exact wavefunction of the bound states
in the QD. To investigate the intervalley coupling in the more general
case of the QDs with noncircular shape, we consider here the alternative
approach of real-space tight-bind (RSTB) method. Using RSTB, we first
calculate intervalley couplings of circular shaped QDs and compare
with the results from EFM. The agreement between the two methods well
justifies the validity of the RSTB method. Then we calculate intervalley
couplings of noncircular shaped QDs and analyze the results. Since
SOC has negligible effects on intervalley coupling, we did not take
into account SOC in the RSTB calculation.

\subsection{The RSTB model}

We use supercell and periodic boundary condition to model a QD. The
supercell is a square with side length $L_{{\rm sc}}$ (cf. figure
\ref{fig:nonCirPot}). Each supercell has $N=L_{{\rm sc}}*{\rm round}(2L_{{\rm sc}}/\sqrt{3})$
lattice points for primitive cells. All lengths are in unit of lattice
constant $a$. We adopt the three-band tight-binding model developed
in  \cite{Liu_Xiao_2013_88_85433__Three}, considering hoppings between
nearest-neighbour Mo-$d_{z^{2}}$, $d_{xy}$, and $d_{x^{2}-y^{2}}$
orbitals. The coordinates of the lattice sites of Mo atoms are denoted
as $\vr_{n}$ $(n=1,2,\cdots,N)$. The tight-binding Hamiltonian of
the QD is

\begin{equation}
H=\sum_{i=1}^{N}\sum_{\alpha}\big[\epsilon_{\alpha}+U(\vr_{i})\big]c_{i\alpha}^{\dagger}c_{i\alpha}+\sum_{\langle i,j\rangle}\sum_{\alpha,\beta}t_{i\alpha,j\beta}c_{i\alpha}^{\dagger}c_{j\beta},
\end{equation}
where $U(\vr_{i})$ is the QD confinement potential at $\vr_{i}$,
$\alpha,\beta=d_{z^{2}},\, d_{xy},\, d_{x^{2}-y^{2}}$ are orbital
indices, $c_{i\alpha}^{\dagger}$ is the creation operator for electron
on site $i$ with orbital $\alpha$. $t_{i\alpha,j\beta}$ is the
hopping between $\alpha$ orbital at position $i$ and $\beta$ orbital
at position $j$, and $\langle,\rangle$ denotes summation over nearest-neighbor
pairs only. In practice, we write the Hamiltonian $H$ in form of
a $3N\times3N$ matrix. The part of $H$ involving hoppings between
site $i$ and $j$ can be written as a $3\times3$ block $H_{ij}=h(\vr_{j}-\vr_{i})$,
where $\vr_{j}-\vr_{i}$ is either $\bm{0}$ or the six nearest-neighbour
vectors $\vR_{1}$ to $\vR_{6}$ defined in  \cite{Liu_Xiao_2013_88_85433__Three}.
Using the real hopping parameters $\{\epsilon_{1},\epsilon_{2},t_{0},t_{1},t_{2},t_{11},t_{12},t_{22}\}$
defined in  \cite{Liu_Xiao_2013_88_85433__Three} (the GGA-version
parameters therein), we have,

\begin{equation}
h(\bm{0})=\begin{bmatrix}\begin{array}{ccc}
\epsilon_{1} & 0 & 0\\
0 & \epsilon_{2} & 0\\
0 & 0 & \epsilon_{2}
\end{array}\end{bmatrix},\ \ \ \ \ \ \ h(\vR_{1})=\begin{bmatrix}\begin{array}{ccc}
t_{0} & t_{1} & t_{2}\\
-t_{1} & t_{11} & t_{12}\\
t_{2} & -t_{12} & t_{22}
\end{array}\end{bmatrix},
\end{equation}

\begin{widetext}

\begin{equation}
h(\vR_{2})=\begin{bmatrix}\begin{array}{ccc}
t_{0} & \frac{1}{2}\left(t_{1}-\sqrt{3}t_{2}\right) & -\frac{1}{2}\left(\sqrt{3}t_{1}+t_{2}\right)\\
-\frac{1}{2}\left(t_{1}+\sqrt{3}t_{2}\right) & \frac{1}{4}\left(t_{11}+3t_{22}\right) & -\frac{\sqrt{3}}{4}(t_{11}-t_{22})-t_{12}\\
\frac{1}{2}\left(\sqrt{3}t_{1}-t_{2}\right) & -\frac{\sqrt{3}}{4}(t_{11}-t_{22})+t_{12} & \frac{1}{4}\left(3t_{11}+t_{22}\right)
\end{array}\end{bmatrix},
\end{equation}

\begin{equation}
h(\vR_{3})=\begin{bmatrix}\begin{array}{ccc}
t_{0} & -\frac{1}{2}\left(t_{1}-\sqrt{3}t_{2}\right) & -\frac{1}{2}\left(\sqrt{3}t_{1}+t_{2}\right)\\
\frac{1}{2}\left(t_{1}+\sqrt{3}t_{2}\right) & \frac{1}{4}\left(t_{11}+3t_{22}\right) & \frac{\sqrt{3}}{4}(t_{11}-t_{22})+t_{12}\\
\frac{1}{2}\left(\sqrt{3}t_{1}-t_{2}\right) & \frac{\sqrt{3}}{4}(t_{11}-t_{22})-t_{12} & \frac{1}{4}\left(3t_{11}+t_{22}\right)
\end{array}\end{bmatrix},
\end{equation}
\end{widetext}

\begin{equation}
h(\vR_{4})=h(\vR_{1})^{\dagger},\ \ \ \ h(\vR_{5})=h(\vR_{2}){}^{\dagger},\ \ \ \ h(\vR_{6})=h(\vR_{3})^{\dagger}.
\end{equation}
The $3N\times3N$ RSTB Hamiltonian matrix for a QD is then diagonalized
to find the eigenstates and eigenenergies. 

Unlike the EFM approach where the bound state wavefunctions are first
given in each valley, and the intervalley coupling then calculated
as the off-diagonal matrix element between them {[}cf. equation \eqref{eq:Vvcs}{]},
the eigenstates and eigenenergies of the RSTB Hamiltonian $H$ already
include the effects of intervalley coupling. The intervalley coupling
causes a fine splitting of the energy levels, which otherwise have
the two-fold valley degeneracy in the absence of SOC. This energy
splitting is just $2|V_{{\rm inter}}|$. Thus, the magnitude of the
intervalley coupling $V_{{\rm inter}}$ is read out from the difference
between the lowest two conduction-band energy eigenvalues of $H$
in the RSTB solution \cite{note1}. 

\begin{figure}
\centering{}\includegraphics[width=8cm]{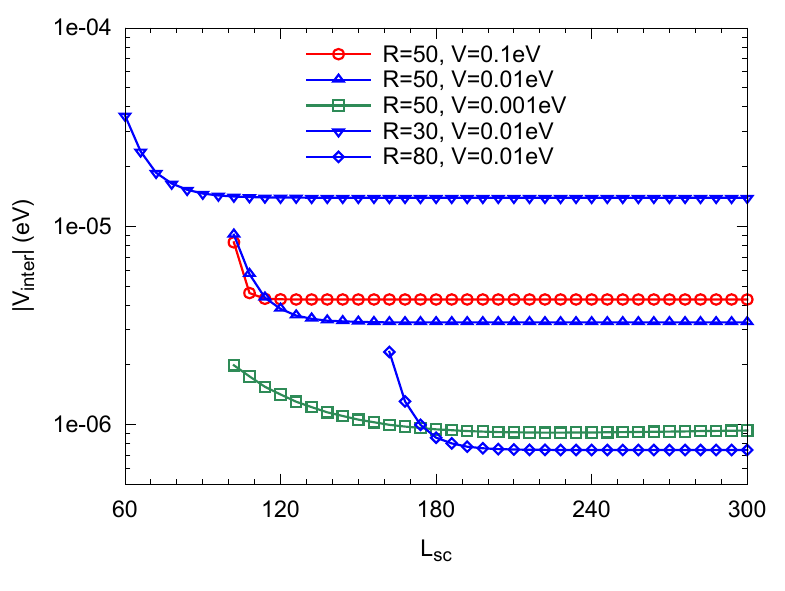}\caption{Intervalley coupling strength $|V_{{\rm inter}}|$ of circular QDs
($L=0$) calculated with RSTB, when the supercell size $L_{{\rm sc}}$
used in the calculation is varied (cf. figure \ref{fig:nonCirPot}).
$|V_{{\rm inter}}|$ converges fast with increasing $L_{{\rm sc}}$.
All lengths are in the unit of $a$. \label{fig:Vvc-Lsc}}
\end{figure}

\subsection{Circular shaped QDs}

We study here circular shaped QDs using RSTB and compare with the
results in EFM. First, we check the impact of supercell size on the
intervalley coupling. Figure \ref{fig:Vvc-Lsc} shows the $|V_{{\rm inter}}|$
vs. $L_{{\rm sc}}$ relation for different circular QDs. We can see
that $|V_{{\rm inter}}|$ converges fast with increasing $L_{{\rm sc}}$,
which means that the finite-size effects introduced by the method
can be well eliminated by having large enough supercell. Figure \ref{fig:Vvc-Lsc}
also shows that $|V_{{\rm inter}}|$ converges faster with increasing
$L_{{\rm sc}}$ for larger potential depth $V$, just as expected. 

\begin{figure}
\centering{}\includegraphics[width=15cm]{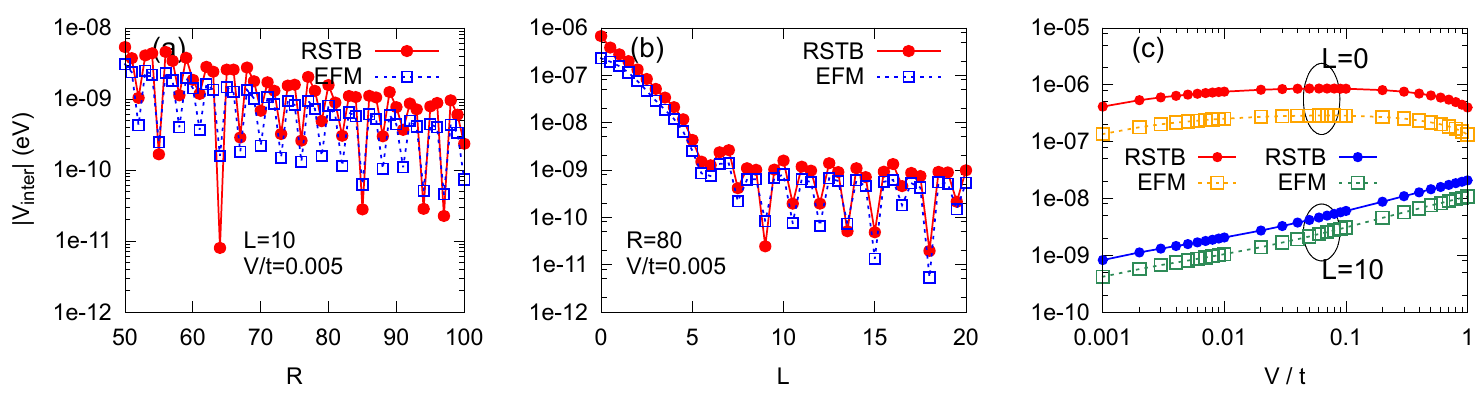}\caption{Comparisons of $|V_{{\rm inter}}|$ of circular QDs calculated by
the RSTB method and the EFM, in the absence of SOC. (a) $|V_{{\rm inter}}|$
vs $R$ ($L=10$, $V/t=0.005$); (b) $|V_{{\rm inter}}|$ vs $L$
($R=80$, $V/t=0.005$); (c) $|V_{{\rm inter}}|$ vs $V$ ($R=80$,
$L=0$ and 10). $L_{{\rm sc}}=240$ for all. All lengths are in the
unit of $a$. \label{fig:Vvc-cmp}}
\end{figure}

$|V_{{\rm inter}}|$ calculated using RSTB method with a large supercell
($L_{{\rm sc}}=240$) is compared with the results from EFM in figure
\ref{fig:Vvc-cmp}. For the $|V_{{\rm inter}}|$ vs. $R$ dependence
{[}figure \ref{fig:Vvc-cmp}(a){]}, the $|V_{{\rm inter}}|$ vs. $L$
dependence {[}figure \ref{fig:Vvc-cmp}(b){]}, and the $|V_{{\rm inter}}|$
vs. $V$ dependence {[}figure \ref{fig:Vvc-cmp}(c){]}, the RSTB data,
including the rapid local oscillations, agree quantitatively well
with the EFM ones. The EFM results rely on the details of the first-principles
wavefunctions of Bloch states from the ABINIT package, while the RSTB
calculations are based on a few hopping matrix elements only. This\textbf{
} agreement between the two entirely different approaches well justify
both the RSTB method and EFM.

\begin{figure}
\centering{}\includegraphics[width=14cm]{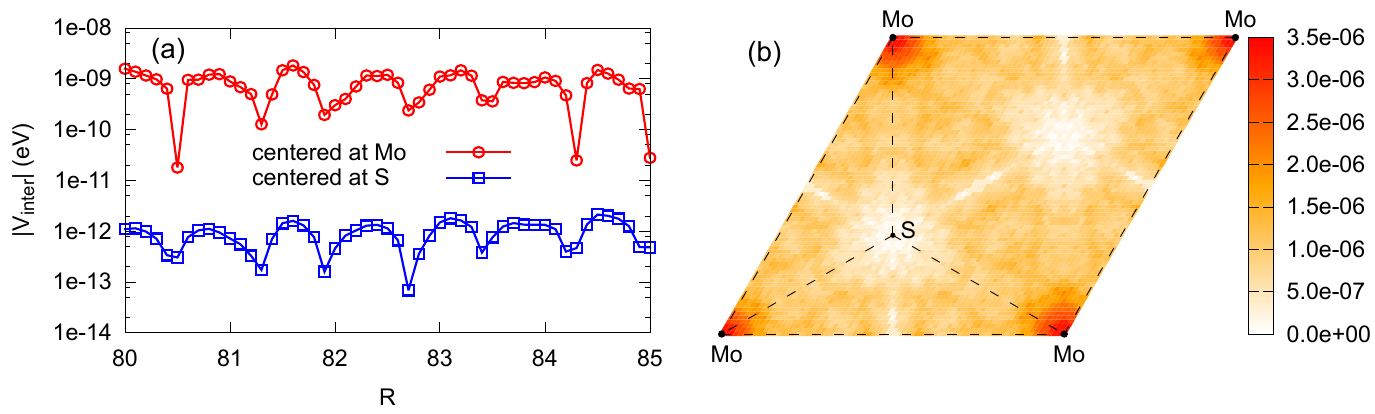}\caption{Dependence of $|V_{{\rm inter}}|$ on the central position of the
circular shaped QD potential, obtained in RSTB. (a) $|V_{{\rm inter}}|$
vs $R$ for the potentials centered at Mo-site (red circle) and S-site
(blue square) respectively. $L=10$, $V=0.0055$eV, $L_{{\rm sc}}=240$.
(b) $|V_{{\rm inter}}|$ as a function of the central position of
the potential in a unit cell. $R=50$, $L=0$, $V=0.01$eV, $L_{{\rm sc}}=180$.
All lengths are in the unit of $a$. \label{fig:Vvc-cell}}
\end{figure}

In figure \ref{fig:Vvc-cell}, we analyze the dependence of the intervalley
coupling strength on the central position of the confinement potential.
Figure \ref{fig:Vvc-cell}(a) shows $|V_{{\rm inter}}|$ vs. $R$
curves of the same circular shaped potential centered at a Mo or S
site. Consistent with the symmetry analysis in Section \ref{sec:Sym},
$|V_{{\rm inter}}|$ is three orders of magnitude larger for the Mo-centered
case compared to the S-centered ate, the latter can be regarded as
the numerical errors from zero. Figure \ref{fig:Vvc-cell}(b) plots
$|V_{{\rm inter}}|$ as a function of the central position of the
confinement potential. Clearly, $|V_{{\rm inter}}|$ decreases when
the potential center moves away from the Mo site and eventually vanishes
when the potential center is at the S site.

\subsection{Intervalley coupling in noncircular QD potentials with vertical wall
($L=0$)}

\begin{figure}
\centering{}\includegraphics[width=15cm]{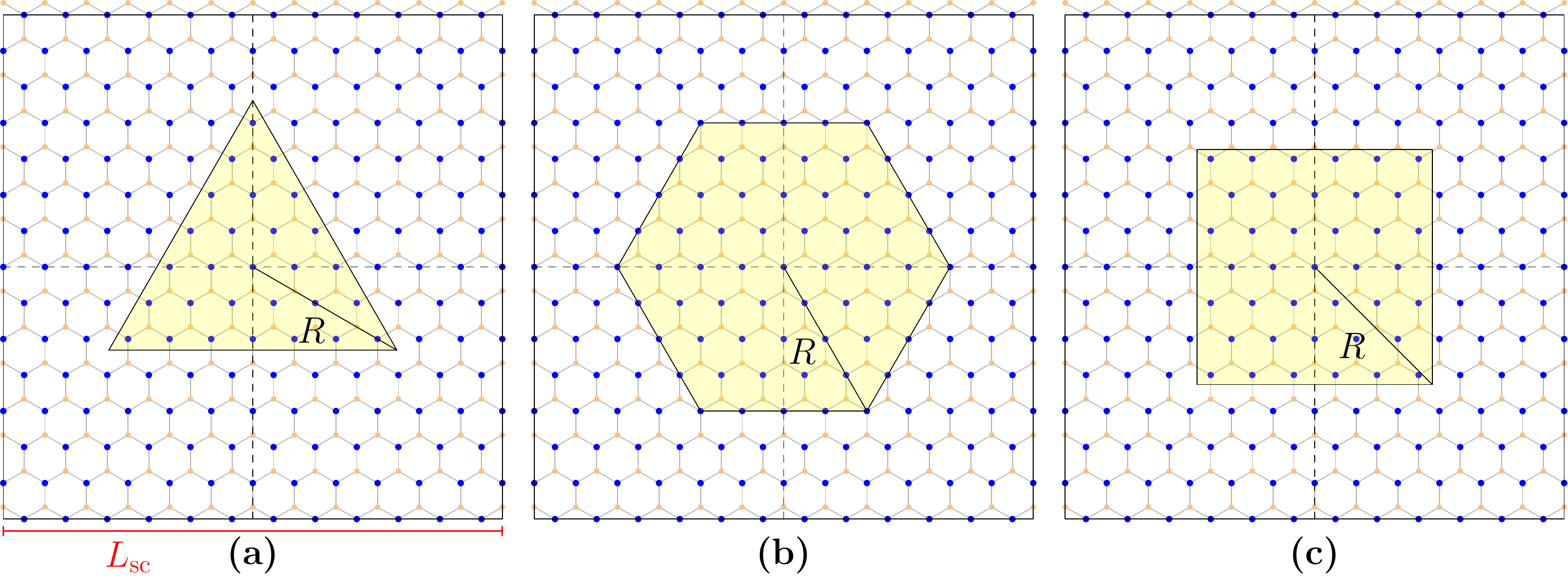}\caption{Schematics of noncircular QD potentials with triangular (a), hexagonal
(b), and square shapes (c). The centers of the QD potentials are at
the Mo sites (blue dots). $R$ is the effective radius. The outer
box illustrates the supercell with size $L_{{\rm sc}}$ ($L_{{\rm sc}}=12$
here). All lengths are in the unit of $a$. \label{fig:nonCirPot}}
\end{figure}

QDs can be formed by the lateral heterostructures between different
monolayer TMDs. Currently such heterostructures are grown by chemical
vapor deposition (CVD), where the shape of confinement is triangular
\cite{Huang_Xu_2014____Lateral}. The lateral size of the confinement
in available heterostructures is several $\mu$m, still too large
for QDs. Nevertheless, growth of monolayer TMDs flakes with sizes
ranging from tens of nm to tens of \textbf{$\mu$}m and with both
triangular and hexagonal shapes have been reported by various groups\textbf{
}\cite{Helveg_Besenbacher_2000_84_951__Atomic,bollinger_one_dimensional_2001,Zande_Hone_2013_12_554__Grains,Peimyoo_Yu_2013_7_10985__Nonblinking,Najmaei_Lou_2013_12_754__Vapour,Cong_Yu_2014_2_131__Synthesis,Wu_Xu_2013_7_2768__Vapor,Gutierrez_Terrones_2013_13_3447__Extraordinary}.
Smaller QDs formed by the lateral heterostructures can be well expected.
In such heterostructures, the confinement potential is due to the
band edge difference between the TMDs. Thus the potential well has
a vertical wall at the QD boundary, corresponding to the $L=0$ limit
discussed above. Here we use RSTB method to investigate intervalley
coupling of these noncircular QDs. We studied QDs with triangular,
hexagonal as well as the square shapes for comparison. 

As shown in figure \ref{fig:nonCirPot}, the triangular, hexagonal,
and square QDs are defined by the confinement potentials $U_{{\rm T}}(\vr)$,
$U_{{\rm H}}(\vr)$, and $U_{{\rm S}}(\vr)$ respectively, where $U_{{\rm T/H/S}}$
equals to 0 within the QDs (yellow regions), and equals to the constant
$V$ outside (white regions). The 'radius' of the QDs is denoted as
$R$ (see figure \ref{fig:nonCirPot}). For the triangular and hexagonal
QDs, we have chosen the orientation of the confinement potential such
that the sides are all along the zigzag crystalline axes, corresponding
to those realizable by the lateral heterostructures of different dichalcogenides.
For the square QDs, two sides are along the zigzag and the others
are along the armchair crystalline axes. 

\begin{figure}
\centering{}\includegraphics[width=15cm]{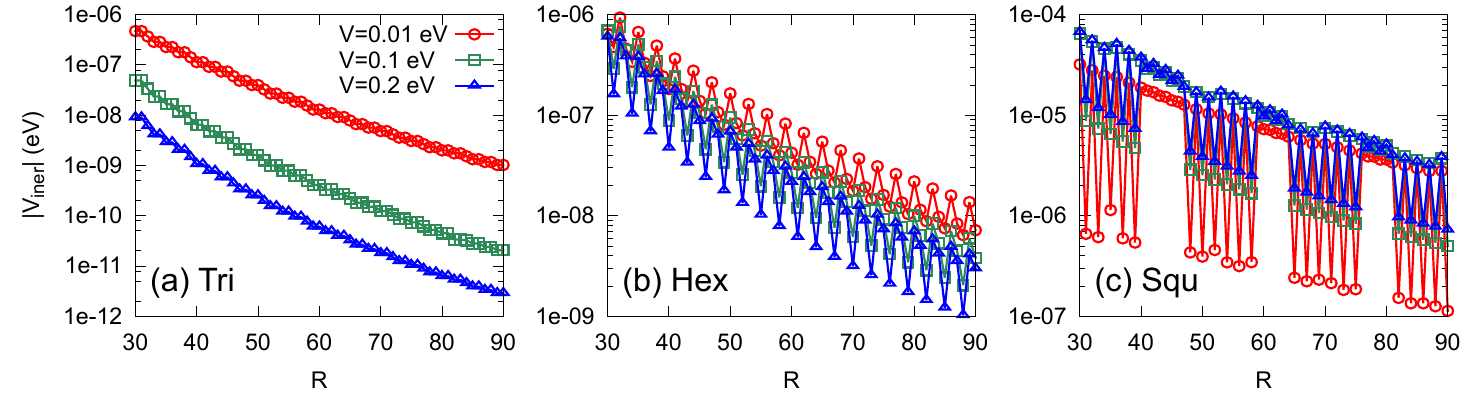}\caption{$|V_{{\rm inter}}|$ vs $R$ relations for triangular QDs (a), hexagonal
QDs (b), and square QDs (c), calculated using RSTB, for $V=0.01$eV
(red circle), 0.1 eV (green square), and 0.2 eV (blue triangle) respectively.
Supercell size is $L_{{\rm sc}}=240$. All lengths are in the unit
of $a$. \label{fig:Vvc-R-THS}}
\end{figure}

\begin{figure}
\centering{}\includegraphics[width=8cm]{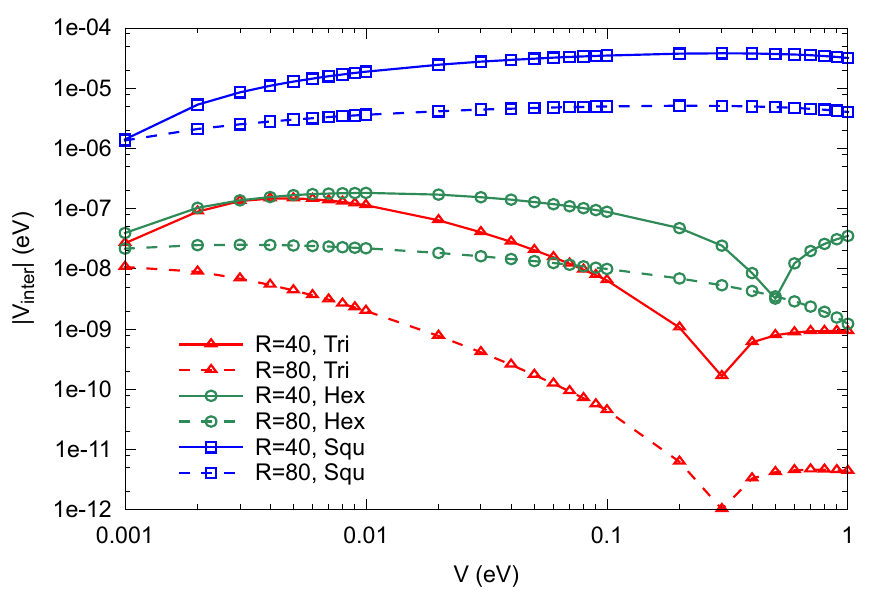}\caption{$|V_{{\rm inter}}|$ vs $V$ relations for triangular (red), hexagonal
(green), and square (blue) QDs, calculated using RSTB. $L_{{\rm sc}}=240$
is used. All lengths are in the unit of $a$. \label{fig:Vvc-V-THS}}
\end{figure}

Figure \ref{fig:Vvc-R-THS} plots the intervalley coupling as function
of the QDs size, for three different potential depths$V=0.01\,$eV,
$V=0.1\,$eV and $V=0.2\,$eV. With increasing $R$, the intervalley
coupling strength has an overall trend to decrease, but with rapid
local oscillations. These features are similar to the circular QDs
(cf. figure \ref{fig:Vvc-R}). Interestingly, for the triangular and
hexagonal QDs, intervalley coupling is found to be much smaller with
the larger potential jump $V$ at the QD boundary. In figure \ref{fig:Vvc-V-THS},
the dependence of the intervalley coupling on $V$ is investigated,
in which all curves are non-monotonic. The non-monotonic dependence
of intervalley coupling strength is also found for the circular QDs
when $L=0$ (cf. figure \ref{fig:Vvc-V}). We note that in heterostructures
formed between different monolayer dichalcogenides (e.g. MoSe$_{2}$
and WSe$_{2}$), the conduction and valence band edge discontinuities
are discovered to be in the range of 0.2--0.4 eV \cite{Rivera_Xu_2014___1403.4985_Observation}.
For this range of values for $V$, the intervalley coupling is well
negligible for the triangular and hexagonal QDs. 

\begin{figure}
\centering{}\includegraphics[width=15cm]{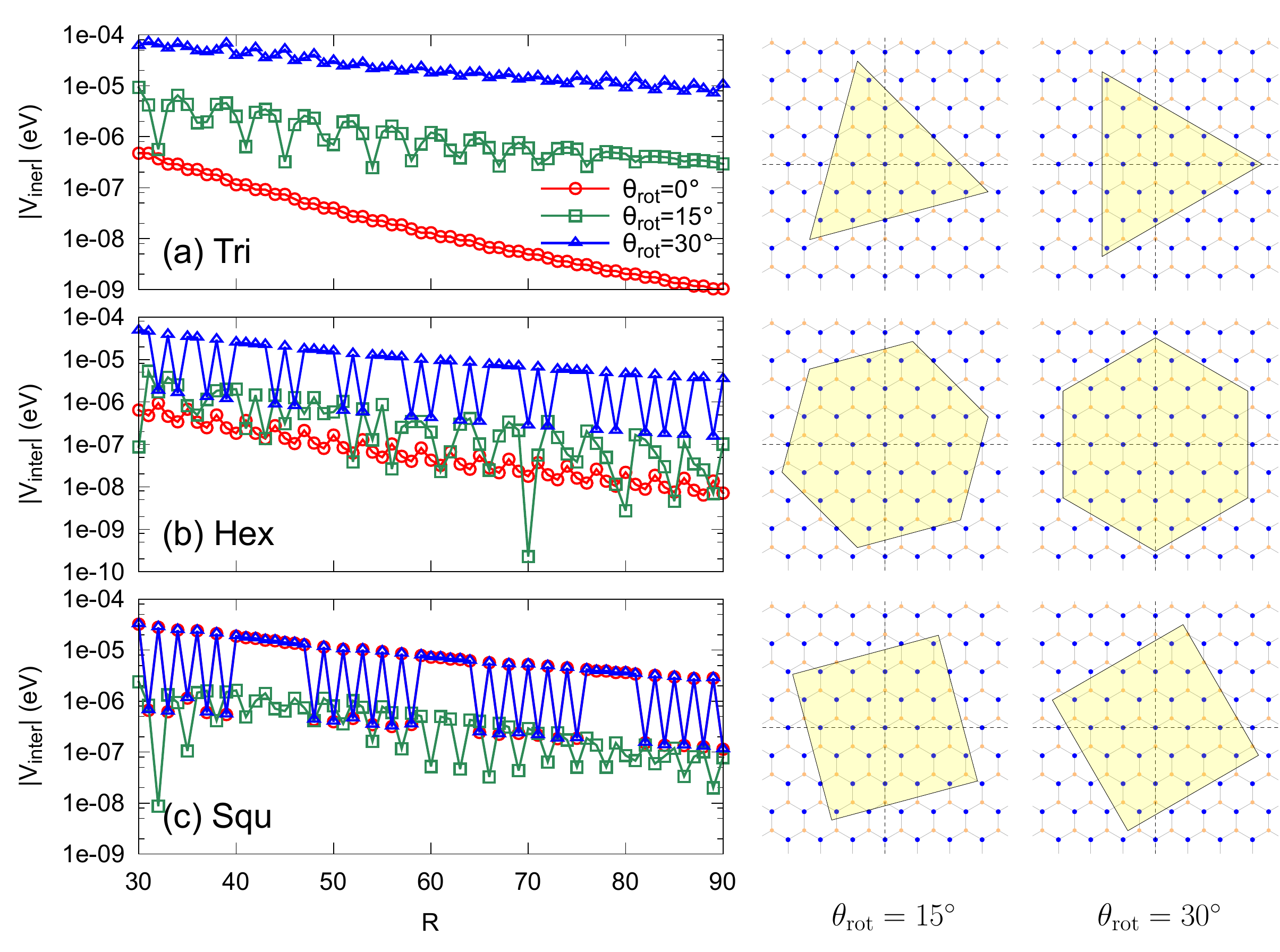}\caption{Size and orientation dependence of $|V_{{\rm inter}}|$ for triangular
QDs (a), hexagonal QDs (b), and square QDs (c), calculated using RSTB.
The orientation of the QD potentials relative to the lattice is defined
by the rotation angle $\theta_{{\rm rot}}$. $\theta_{{\rm rot}}=0^{\circ}$
corresponds to the configurations shown in figure \ref{fig:nonCirPot}.
The orientations with $\theta_{{\rm rot}}=15^{\circ}$ and $\theta_{{\rm rot}}=30^{\circ}$
are shown in the right panels here. For QDs of each shape, $|V_{{\rm inter}}|$
vs $R$ relations are compared for the orientations $\theta_{{\rm rot}}=0^{\circ}$,
$\theta_{{\rm rot}}=15^{\circ}$, and $\theta_{{\rm rot}}=30^{\circ}$.
For square QDs, $\theta_{{\rm rot}}=0^{\circ}$ and $\theta_{{\rm rot}}=30^{\circ}$
correspond to the same configuration. $V=0.01$eV and $L_{{\rm sc}}=240$.
All lengths are in the unit of $a$. \label{fig:Vvc-R-THS-rot}}
\end{figure}

Comparing QDs of different shapes, the intervalley coupling in square
QDs is orders of magnitude larger than that in the triangular and
hexagonal QDs. Such a sharp difference is due to the relative orientation
of the sides of the confinement potential to the crystalline axes.
Inhomogeneous junctions along the zigzag (armchair) crystalline axes
preserves the momentum along (perpendicular to) the lines connecting
the two valleys in the momentum space. Thus, the zigzag junctions
introduce the minimum intervalley coupling, while the armchair junctions
cause maximum intervalley coupling. For the triangular and hexagonal
QDs, all sides are along the zigzag crystalline axes, while for square
QDs two sides are along the armchair directions and thus intervalley
coupling is much larger. This is further verified by examine the change
of the intervalley coupling when we rotate the QD confinement potential
by an angle $\theta_{{\rm rot}}$ relative to the orientation defined
in figure \ref{fig:nonCirPot}, as shown in figure \ref{fig:Vvc-R-THS-rot}.
For triangular and hexagonal QDs, we find the intervalley coupling
increases with $\theta_{{\rm rot}}$ significantly, and reaches a
maximum at $\theta_{{\rm rot}}=30^{\circ}$ where the intervalley
coupling becomes comparable with the square QDs. Note that at $\theta_{{\rm rot}}=30^{\circ}$,
the sides of the confinement potential of the triangular and hexagonal
QDs are along the armchair crystalline axes {[}cf. figure \ref{fig:Vvc-R-THS-rot}(a)
and (b){]}. 

\begin{figure}
\centering{}\includegraphics[width=12cm]{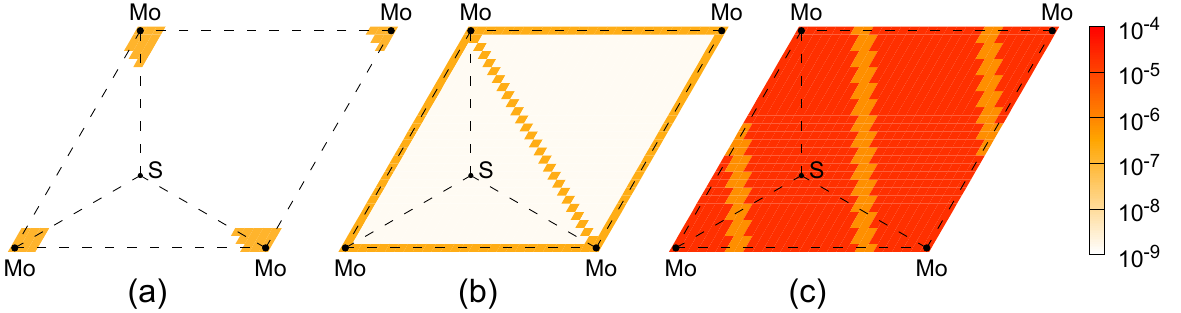}\caption{$|V_{{\rm inter}}|$ as a function of the central position of the
potential in a unit cell for triangular QDs (a), hexagonal QDs (b),
and square QDs (c), calculated using RSTB. $R=40$, $V=0.01$eV, and
$L_{{\rm sc}}=120$ are used. Resolution of the maps is 27$\times$27.
All lengths are in the unit of $a$, and the unit of energy is eV.
\label{fig:cell-THS}}
\end{figure}

Figure \ref{fig:cell-THS} shows the dependence of $|V_{{\rm inter}}|$
on the central position of the confinement potential in a unit cell
for the three kinds of noncircular QDs. It is clear that $|V_{{\rm inter}}|$
strongly depends on the potential center. For triangular and hexagonal
QDs with the $C_{3}$ rotational symmetry, the dependence is similar
to the circular QDs {[}cf. figure \ref{fig:Vvc-cell}(b){]}, i.e.
intervalley coupling is strongest if the potential is centered at
a Mo site, but vanishes if the potential is centered at a S site.
Such behavior is absent for the square QDs that lacks the $C_{3}$
rotational symmetry.

\subsection{Intervalley coupling in smooth noncircular QD potential}

\begin{figure}
\centering{}\includegraphics[width=14cm]{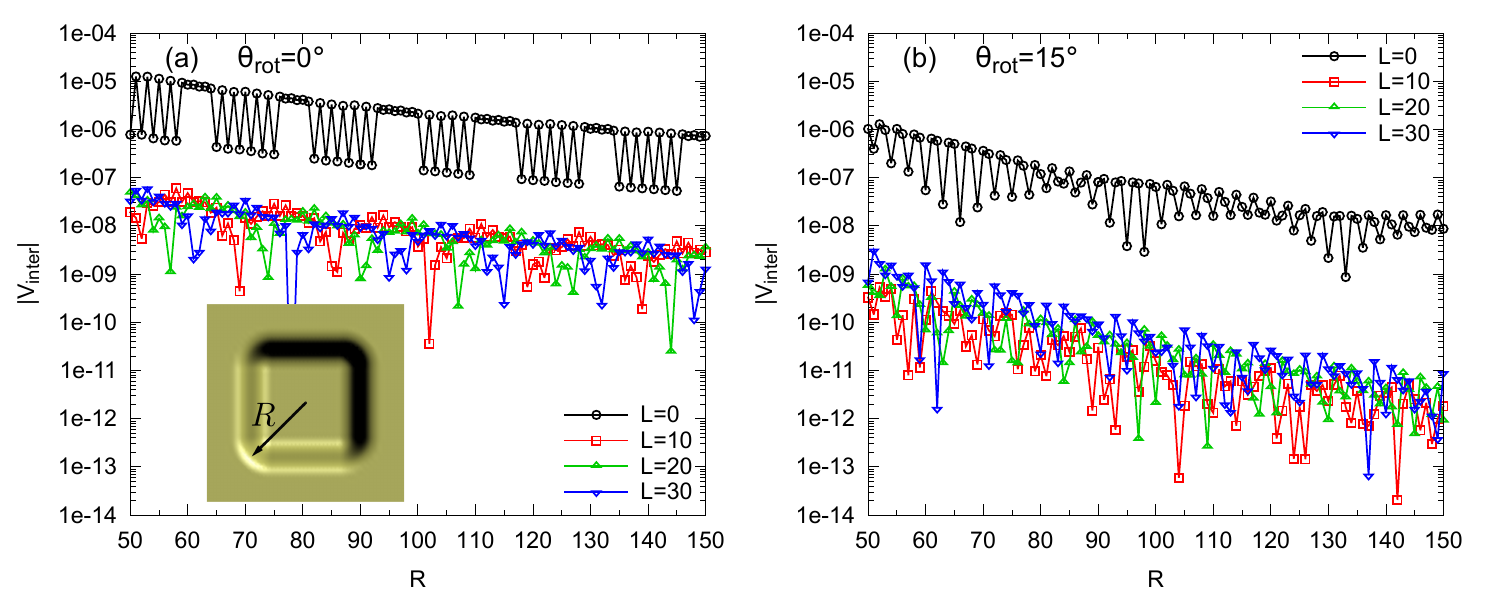}\caption{Intervalley coupling strength for square-shaped QDs. Potentials with
smooth walls {[}finite $L$, see the inset of (a) for the schematic
illustration{]} of different slops are compared with the one with
vertical wall. (a) and (b) for the orientations $\theta_{{\rm rot}}=0^{\circ}$
and $\theta_{{\rm rot}}=15^{\circ}$ respectively. Red square symbol
is for $L=$10, green upward triangle for $L=$20, blue downward triangle
for $L=$ 30, and black circle for $L=0$. $V=0.02$ eV. $L_{{\rm sc}}=240$
for $R\in[50,90]$ and $L_{{\rm sc}}=360$ for $R\in[90,150]$. All
lengths are in the unit of $a$, and the unit of energy is eV. \label{fig:VvcR-RndSqu}}
\end{figure}

Finally, we examine square shaped QDs but with a smooth slope for
the potential change on the boundary (i.e. finite $L$), which represents
a general situation for QDs defined by patterned electrodes. The calculated
intervalley coupling strength $|V_{{\rm inter}}|$ as a function of
QD size $R$ is shown in figure \ref{fig:VvcR-RndSqu}. The peak value
of intervalley coupling decreases nearly exponentially with the QD
size. The dependence on the smoothness $L$ is similar to the circular
shaped QDs (cf. figure \ref{fig:Vvc-L}), i.e. $|V_{{\rm inter}}|$
drops fast by several orders of magnitude when $L$ increases from
0 to 5, and becomes largely independent of $L$ for $L>0$. From figure
\ref{fig:VvcR-RndSqu}, we can see that the $|V_{{\rm inter}}|$ curves
as functions of $R$ for $L=10$, $L=20$, and $L=30$ are basically
the same, except for the location of some fine dips.

\section{conclusions and discussions\label{sec:Conclusions}}

To conclude, we have investigated intervalley coupling in QDs defined
by confinement potentials of various shapes on extended TMD monolayer.
The numerical results obtained using two completely different approaches
agree well with each other. For confinement potentials with the $C_{3}$
rotational symmetry, the intervalley coupling is maximized (zero)
if the potential is centered at a M (X) site. Comparing smooth confinement
potentials with slopping walls and sharp confinement potentials with
vertical walls, the intervalley coupling in the latter is much stronger.
When the length scales of the QDs vary, the intervalley coupling exhibits
fast oscillations where the maxima can be two orders of magnitude
larger than the neighboring minima, and the envelop of the maxima
has an overall nearly exponential decay with the increase of the QDs
size. When all parameters are comparable, the intervalley coupling
can be smaller by several orders of magnitude in circular shaped QDs,
and in triangular and hexagonal shaped QDs with all sides along the
zigzag crystalline axes. For all QDs studied, the largest intervalley
coupling is upper bounded by 0.1 meV, found for small QDs with diameters
of 20 nm and with sharp confinement potentials (i.e. vertical walls). 

The intervalley coupling in these monolayer TMD QDs is much smaller
compared to that in graphene nanoribbon QDs \cite{Trauzettel_Burkard_2007_3_192__Spin}
and in silicon QDs \cite{Saraiva_Koiller_2009_80_81305__Physical,Koiller_Das_2004_70_115207__Shallow,Goswami_Eriksson_2007_3_41__Controllable,Friesen_Coppersmith_2010_81_115324__Theory,Culcer_Das_2010_82_155312__Quantum,Friesen_Coppersmith_2007_75_115318__Valley,Nestoklon_Ivchenko_2006_73_235334__Spin,Boykin_Lee_2004_84_115__Valley}.
In QDs formed in 2D crystals, intervalley coupling mainly occurs at
the boundary where the translational invariance is lost, thus the
coupling strength is proportional to the probability distribution
of the electron at the QD boundary. In the TMD QDs studied here, the
electron wavefunction has vanishing amplitude at the QD boundary {[}cf.
green shaded region in figure \ref{fig:ene}(b){]}. This is a characteristic
of the QDs defined by confinement potentials on an otherwise extended
crystalline monolayer, as opposed to the nanoribbon QDs \cite{Trauzettel_Burkard_2007_3_192__Spin}.
The geometry here is also qualitatively different from the QDs formed
in silicon inversion layers. In those silicon QDs, the two valleys
are separated by a wavevector along $z$-direction (perpendicular
to the inversion layer), and their coupling is caused by the sharp
confinement with a length scale of nm in the $z$-direction \cite{Saraiva_Koiller_2009_80_81305__Physical,Koiller_Das_2004_70_115207__Shallow,Goswami_Eriksson_2007_3_41__Controllable,Friesen_Coppersmith_2010_81_115324__Theory,Culcer_Das_2010_82_155312__Quantum,Friesen_Coppersmith_2007_75_115318__Valley,Nestoklon_Ivchenko_2006_73_235334__Spin,Boykin_Lee_2004_84_115__Valley}.
Such confinement strongly squeezes the wavefunction, giving rise to
its large amplitude at the interface where valley hybridization occurs,
and hence there is a stronger intervalley coupling in the order of
meV. In contrast, here it is the much larger lateral size $R$ of
the QDs that determines the wavefunction amplitude at the QD boundary.
We can expect that intervalley coupling is generically small in QDs
defined by confinement potentials on otherwise extended monolayer.

Because of the very small intervalley coupling in the monolayer TMD
QDs, valley hybridization is then well quenched by the much stronger
spin-valley coupling of the electron in monolayer TMDs. As a result,
the QDs can well inherit the valley physics of the 2D bulk such as
the valley optical selection rules, making the valley pseudospin of
single confined electron a perfect candidate of quantum bit. The sensitive
dependence of intervalley coupling strength on the central position
and the lateral length scales of the confinement potentials further
makes possible the tuning of the intervalley coupling by external
controls. The qualitative conclusions here are also applicable to
QDs formed by confinement potentials on other 2D materials with finite
gap. 
\begin{acknowledgments}
The work was supported by the Croucher Foundation under the Croucher
Innovation Award, the Research Grant Council of HKSAR under Grant
No. HKU705513P and HKU9/CRF/13G (G.B.L., H.P., and W.Y.); the NSFC
with Grant No. 11304014, the National Basic Research Program of China
973 Program with Grant No. 2013CB934500 and the Basic Research Funds
of Beijing Institute of Technology with Grant No. 20131842001 and
20121842003 (G.B.L.); the MOST Project of China with Grants Nos. 2014CB920903
and 2011CBA00100, the NSFC with Grant Nos. 11174337 and 11225418,
the Specialized Research Fund for the Doctoral Program of Higher Education
of China with Grants No. 20121101110046 (Y.Y.).
\end{acknowledgments}
\appendix

\section{The eigenvalue of Bloch function under $C_{3}$ rotation\label{sec:C3rot}}

Here we give a derivation of $\gamma_{{\rm \alpha}}^{\tau}$,
the eigenvalue of the Bloch function $\varphi_{\alpha}^{\tau}$ under
the $C_{3}$ rotation. We first express $\varphi_{\alpha}^{\tau}$
in terms of linear combination of atomic orbitals $d_{\alpha}$
($\alpha=$c, v),
\begin{equation}
\varphi_{\alpha}^{\tau}(\vr)=\frac{1}{\sqrt{N}}\sum_{\vR}e^{i\tau K\cdot(\vR+\bm{\delta})}d_{\alpha}(\vr-\vR-\bm{\delta}),
\end{equation}
in which $\vR$ is lattice vector, $\bm{\delta}$ is the position
of Mo atom in the unit cell, $d_{{\rm c}}=d_{0}=d_{z^{2}}$, and $d_{{\rm v}}=d_{\tau2}=\frac{1}{\sqrt{2}}(d_{x^{2}-y^{2}}+i\tau d_{xy})$.
According the definition of $\gamma_{\alpha}^{\tau}\equiv[C_{3}\varphi_{\alpha}^{\tau}(\vr)]/\varphi_{\alpha}^{\tau}(\vr)$
and using the orthogonality of $C_{3}$ ($\vk\cdot\vr=C_{3}\vk\cdot C_{3}\vr$),
we have
\begin{eqnarray}
C_{3}\varphi_{\alpha}^{\tau}(\vr) & = & \frac{1}{\sqrt{N}}\sum_{\vR}e^{i\tau K\cdot(\vR+\bm{\delta})}d_{\alpha}(C_{3}^{-1}\vr-\vR-\bm{\delta})=\frac{1}{\sqrt{N}}\sum_{\vR}e^{i\tau C_{3}K\cdot C_{3}(\vR+\bm{\delta})}d_{\alpha}(C_{3}^{-1}(\vr-C_{3}(\vR+\bm{\delta}))).
\end{eqnarray}
Summing over all Mo positions $\{\vR+\bm{\delta}\}$ is equivalent
to summing over all positions $\{C_{3}(\vR+\bm{\delta})\}$ when the rotation
center locates at either Mo or S site (see figure \ref{fig:well}). Accordingly,
with the substitution $C_{3}(\vR+\bm{\delta})\rightarrow(\vR+\bm{\delta})$,
the above equation becomes
\begin{equation}
C_{3}\varphi_{\alpha}^{\tau}(\vr)=\frac{1}{\sqrt{N}}\sum_{\vR}e^{i\tau C_{3}K\cdot(\vR+\bm{\delta})}d_{\alpha}(C_{3}^{-1}(\vr-\vR-\bm{\delta})).\label{eq:C3phi1}
\end{equation}
Because $C_{3}$ is an element of the wave-vector group of $\tau K$,
we have $e^{i\tau C_{3}K\cdot\vR}=e^{i\tau K\cdot\vR}$ and consequently
\begin{equation}
e^{i\tau C_{3}K\cdot(\vR+\bm{\delta})}=e^{i\tau K\cdot\vR}e^{i\tau C_{3}K\cdot\bm{\delta}}=e^{i\tau K\cdot\vR}e^{i\tau K\cdot C_{3}^{-1}\bm{\delta}}=e^{i\tau K\cdot(C_{3}^{-1}\bm{\delta}-\bm{\delta})}e^{i\tau K\cdot(\vR+\bm{\delta})}.\label{eq:etoC3K}
\end{equation}
Define $\gamma_{d_{\alpha}}\equiv[C_{3}d_{\alpha}(\vr)]/d_{\alpha}(\vr)=d_{\alpha}(C_{3}^{-1}\vr)/d_{\alpha}(\vr)$, we have 
\begin{equation}
d_{\alpha}(C_{3}^{-1}(\vr-\vR-\bm{\delta}))=\gamma_{d_{\alpha}}d_{\alpha}(\vr-\vR-\bm{\delta}).\label{eq:gda}
\end{equation}
Plugging equations \eqref{eq:etoC3K} and \eqref{eq:gda} into \eqref{eq:C3phi1},
we obtain
\begin{equation}
C_{3}\varphi_{\alpha}^{\tau}(\vr)=\frac{1}{\sqrt{N}}\sum_{\vR}e^{i\tau K\cdot(C_{3}^{-1}\bm{\delta}-\bm{\delta})}e^{i\tau K\cdot(\vR+\bm{\delta})}\gamma_{d_{\alpha}}d_{\alpha}(\vr-\vR-\bm{\delta})=e^{i\tau K\cdot(C_{3}^{-1}\bm{\delta}-\bm{\delta})}\gamma_{d_{\alpha}}\varphi_{\alpha}^{\tau}(\vr)
\end{equation}
and therefore
\begin{equation}
\gamma_{\alpha}^{\tau}=e^{i\tau K\cdot(C_{3}^{-1}\bm{\delta}-\bm{\delta})}\gamma_{d_{\alpha}},\label{eq:gat}
\end{equation}
in which the factor $\gamma_{d_{\alpha}}$ comes from the rotation of
atomic orbital around its own center and the factor $e^{i\tau K\cdot(C_{3}^{-1}\bm{\delta}-\bm{\delta})}$
comes from the change of the lattice phase of Mo atom after rotation.
According to equation \eqref{eq:gat}, when the rotation center locates
at Mo {[}figure \ref{fig:well}(a){]}, we have $\bm{\delta}=0$ and
hence $\gamma_{\alpha}^{\tau}=\gamma_{d_{\alpha}}$ {[}equation \eqref{eq:C3Mo}{]};
when the rotation center locates at S {[}figure \ref{fig:well}(b){]},
we have $\bm{\delta}=(1,\frac{1}{\sqrt{3}})a$ and hence $\gamma_{\alpha}^{\tau}=e^{i\tau\frac{2\pi}{3}}\gamma_{d_{\alpha}}$
{[}equation \eqref{eq:C3S}{]}.

\bibliographystyle{apsrev4-1}
\bibliography{refs}

\end{document}